\documentclass[11pt]{article}
\pdfoutput=1  
\usepackage{graphicx}
\graphicspath{{./figures/}}
\usepackage{appendix}
\usepackage{latexsym,amsmath,amsfonts,amssymb,booktabs}
\usepackage[font=small]{caption}
\usepackage{slashed,upgreek,amscd,cancel,tensor,color}
\usepackage{adjustbox}
\usepackage[numbers,compress,square]{natbib}
\usepackage{epsfig,latexsym}
\usepackage[pdfencoding=auto]{hyperref}
\usepackage{url}
\numberwithin{equation}{section}
\usepackage{doi}
\usepackage{subcaption}
\definecolor{MyBlue}{rgb}{0.15,0.15,0.70}

\hypersetup{
colorlinks=true,
citecolor=MyBlue,
linkcolor=MyBlue,
urlcolor=MyBlue
}

\input epsf
\usepackage{amssymb}
\usepackage{amsmath}
\usepackage{amsfonts}
\usepackage{upgreek}
\usepackage{latexsym}
\usepackage{stfloats}
\usepackage{afterpage}

\newcommand{\be}{\begin{equation}}
\newcommand{\ee}{\end{equation}}
\newcommand{\beq}{\begin{equation}}
\newcommand{\eeq}{\end{equation}}
\newcommand{\bea}{\begin{eqnarray}}
\newcommand{\eea}{\end{eqnarray}}

\newcommand{\R}{R}

\usepackage[stable]{footmisc}

\def\d{\delta}

\def\MM{M_{*}}

\def\dkmu2{\delta K_{\mu \nu}\delta K^{\mu \nu}}
\def\pmu2{  \phi_{\mu \nu}\phi^{\mu \nu}}

\newcommand{\Atwo}{{A_2}}
\newcommand{\Athree}{{A_3}}
\newcommand{\Afour}{{B_4}}
\newcommand{\Afive}{{B_5}}
\newcommand{\Bfour}{{A_4}}
\newcommand{\Bfive}{{A_5}}

\newcommand\bone{{\alpha}_{\rm V1}}
\newcommand\btwo{{\alpha}_{\rm V2}}
\newcommand\bthree{{\alpha}_{\rm V3}}
\newcommand\hatbone{\hat \alpha_{\rm V1} }
\newcommand\hatbtwo{\hat \alpha_{\rm V2} }
\newcommand\hatbthree{\hat \alpha_{\rm V3} }
\newcommand\dotbone{\dot \alpha_{\rm V1} }
\newcommand\dotbtwo{\dot \alpha_{\rm V2} }
\newcommand\dotbthree{\dot \alpha_{\rm V3} }

\newcommand\alphaq{{\alpha}_{\rm V1}}
\newcommand\alphac{{\alpha}_{\rm V2}}

\newcommand\alphaset{{\alpha}_{\rm V3}}
\newcommand\PP{{\cal P}}

\newcommand{\Inv}{A^{-1}}
\newcommand{\deltac}{\delta}

\renewcommand\[{\left[}

\newcommand\ees{\end{eqnarray}}
\newcommand\bees{\begin{eqnarray}}

\newcommand\alphaD{\alpha_{\text{D}}}
\newcommand\alphaC{\alpha_{\text{C}}}

\newcommand\alphaB{\alpha_{\text{B}}}
\newcommand\alphaM{\alpha_{\text{M}}}
\newcommand\alphaK{\alpha_{\text{K}}}
\newcommand\alphaT{\alpha_{\text{T}}}

\newcommand{\eqn}[1]{eq.~(\ref{#1})}

 \newcommand{\half}{\frac{1}{2}}
 \newcommand{\muphi}{\mu_{\Phi}}
 \newcommand{\mupsi}{\mu_{\Psi}}
 \newcommand{\muchi}{\mu_{\chi}}

\newcommand{\omegam}{\Omega_{\rm m}}

\newcommand{\xvec}{\vec{x}}
\newcommand{\knl}{k_{\rm NL}}
\newcommand{\deltam}{\delta}

\newcommand{\barknl}{k_{\rm NL}}
\newcommand{\kV}{k_{V}}

\newcommand{\LambdaU}{\Lambda_{\rm U}} 
\newcommand{\gzz}{ g^{00}}
\newcommand{\dgzz}{\delta g^{00}}

\newcommand{\gtwozt}{G_2^{(0,2)}}
\newcommand{\gtwozf}{G_2^{(0,4)}}

\newcommand{\gtwozth}{G_2^{(0,3)}}

\newcommand{\gthreezo}{G_3^{(0,1)}}
\newcommand{\gthreeoo}{G_3^{(1,1)}}
\newcommand{\gthreezt}{G_3^{(0,2)}}
\newcommand{\gthreeot}{G_3^{(1,2)}}

\newcommand{\gthreezth}{G_3^{(0,3)}}
\newcommand{\gthreezf}{G_3^{(0,4)}}
\newcommand{\gthreeoth}{G_3^{(1,3)}}

\newcommand{\gfouroz}{G_4^{(1,0)}}
\newcommand{\gfourzo}{G_4^{(0,1)}}
\newcommand{\gfouroo}{G_4^{(1,1)}}
\newcommand{\gfourof}{G_4^{(1,4)}}
\newcommand{\gfourzt}{G_4^{(0,2)}}
\newcommand{\gfourzf}{G_4^{(0,4)}}
\newcommand{\gfourot}{G_4^{(1,2)}}
\newcommand{\gfourtz}{G_4^{(2,0)}}

\newcommand{\gfourzth}{G_4^{(0,3)}}
\newcommand{\gfouroth}{G_4^{(1,3)}}

\newcommand{\gfiveoz}{G_5^{(1,0)}}
\newcommand{\gfivezo}{G_5^{(0,1)}}
\newcommand{\gfiveoo}{G_5^{(1,1)}}
\newcommand{\gfivezt}{G_5^{(0,2)}}
\newcommand{\gfivezf}{G_5^{(0,4)}}
\newcommand{\gfiveot}{G_5^{(1,2)}}
\newcommand{\gfivetz}{G_5^{(2,0)}}

\newcommand{\gfivezth}{G_5^{(0,3)}}
\newcommand{\gfiveoth}{G_5^{(1,3)}}

\begin{document}
\vspace{0.5cm}

\begin{center}
\Large{\textbf{Nonlinear Effective Theory of Dark Energy }} \\[1cm]

\large{Giulia Cusin$^a$, Matthew Lewandowski$^b$ and Filippo Vernizzi$^b$}
\\[0.5cm]

\small{
\textit{$^a$ D\'epartement de Physique Th\'eorique and Center for Astroparticle Physics, \\
Universit\'e de Gen\`eve, 24 quai Ansermet, CH--1211 Gen\`eve 4, Switzerland}}

\vspace{.2cm}

\small{
\textit{$^b$ Institut de physique th\' eorique, Universit\'e  Paris Saclay \\ [0.05cm]
CEA, CNRS, 91191 Gif-sur-Yvette, France  }}

\vspace{.2cm}

\vspace{0.5cm}
\today

\end{center}

\vspace{2cm}

\begin{abstract}

We develop an approach to parametrize cosmological perturbations beyond linear order for general dark energy and modified gravity models characterized by a single scalar degree of freedom.  We derive the full nonlinear action, focusing on Horndeski theories. In the quasi-static, non-relativistic limit,
there are a total of six independent relevant operators, three of which start at nonlinear order. The new nonlinear couplings modify, beyond linear order, the generalized Poisson equation relating the Newtonian potential to the matter density contrast. We derive this equation up to cubic order in perturbations and,  in a companion article \cite{CLV2}, we apply it to compute the one-loop matter power spectrum. Within this approach, we also discuss the Vainshtein regime around spherical sources and the relation between the Vainshtein scale and the nonlinear scale for structure formation.
\end{abstract}

\newpage

\tableofcontents

\vspace{.5cm}
\newpage

\section{Introduction}

One of the main goals of future Large-Scale Structure (LSS) surveys will  be to accurately measure the expansion  history of the universe and the evolution of clustering of galaxies and dark matter as a function of redshift. 
Given that gravity has been so far  poorly studied on cosmological scales, these measurements  could open a window on new physics. Indeed, dark energy, the mysterious component responsible for the accelerated expansion of the universe, could affect both the expansion history and the growth of structures (see e.g.~\cite{Amendola:2012ys,Amendola:2016saw} and references therein).

Initially implemented for inflation in \cite{Creminelli:2006xe, Cheung:2007st}, a convenient way to describe dark energy and modified gravity models characterized by a single scalar degree of freedom and to connect them to observational predictions  in terms of a minimal number of parameters 
 is the Effective Field Theory of Dark Energy (EFTofDE)  \cite{Creminelli:2008wc,Gubitosi:2012hu,Bloomfield:2012ff,Gleyzes:2013ooa,Bloomfield:2013efa} (see \cite{Piazza:2013coa,Tsujikawa:2014mba,Gleyzes:2014rba} for reviews); see also \cite{Baker:2011jy,Baker:2012zs,Battye:2012eu,Battye:2013ida} for analogous approaches in dark energy. In the EFTofDE approach, 
one assumes that the time-diffeomorphism invariance of the gravitational sector is broken by the dark-energy field. In the so-called unitary gauge, the gravitational action can be constructed as the sum of all possible operators in terms of the metric, invariant under time-dependent spatial diffeomorphisms and ordered in the number of perturbations and derivatives. Physical principles such as  locality, causality,   stability,  and unitarity can be  imposed at the level of the Lagrangian, so that the predicted signatures are physically acceptable. 

This approach, sometimes combined with the dimensionless parametrization introduced in \cite{Bellini:2014fua} (see \cite{Gleyzes:2014rba}  for the relation between this parametrization and the one used in the EFT approach), has been for instance efficiently applied to derive observational constraints \cite{Ade:2015rim,Bellini:2015xja,Salvatelli:2016mgy}, to study predictions and forecasts \cite{Piazza:2013pua,Perenon:2015sla,Gleyzes:2015rua,Renk:2016olm,Leung:2016xli,Pogosian:2016pwr,DAmico:2016ntq,Alonso:2016suf,Bellomo:2016xhl,Raveri:2017qvt,Gleyzes:2017kpi},
and to develop linear Einstein-Boltzmann codes \cite{Hu:2013twa,Bellini:2015xja,Huang:2012mt}
that can be employed  to compute  standard linear observables, such as the Cosmic Microwave Background (CMB) temperature and polarization anisotropies and the dark matter power spectrum. These codes have been shown to  agree well with each other, up to sub-percent level, for a wide range of wavenumbers \cite{Bellini:2017avd}.

Apart  from some exceptions (see e.g.~\cite{Brax:2015pka,Frusciante:2017nfr}),   the EFTofDE has been developed to include  quadratic operators, sufficient to describe dark energy and modified gravity models in the linear regime, which is applicable on scales above $\sim 10$~Mpc.  However, most of the data coming from future surveys will be on shorter scales, where nonlinearities in matter fluctuations can not be neglected.  On these scales, dark matter density perturbations become large and can feed the dark energy fluctuations.  The quadratic EFTofDE operators are insufficient to describe the dynamics of cosmological perturbations, and higher-order operators must be included.

This work is the first of two papers.  The goal of the present article is to use the EFTofDE to systematically determine the relevant linear and nonlinear operators that are important in the regime of dark-matter clustering.  Then, in \cite{CLV2}, we use these results to compute the one-loop dark-matter power spectrum in the presence of the dark-energy operators studied here.  In this paper, we determine the relevant operators in Sec.~\ref{eftofdesec} (details can be found in App.~\ref{app:Horndeski}) where we also discuss the quasi-static limit and the transformation of the action under change of frame. 
To decide which are the relevant operators, we make two simplifying assumptions. 
First, we restrict the EFT action to describe theories within the Horndeski class \cite{Horndeski:1974wa,Deffayet:2011gz}, which includes all Lorentz-invariant scalar-tensor theories with at most two derivatives in the field equations, and we focus on theories that exhibit Vainshtein screening \cite{Vainshtein:1972sx,Babichev:2013usa}. The radiative stability of these theories has been studied in \cite{Pirtskhalava:2015nla}. For a discussion on theories that are capable of exhibiting more than one type of screening see e.g.~\cite{Gratia:2016tgq}. 

Second, we consider scales much smaller than both the Hubble scale and the sound-horizon of the scalar fluctuations. On these scales, and in the presence of non-relativistic sources, the scalar fluctuations satisfy the quasi-static approximation (see e.g.~\cite{Sawicki:2015zya}). 
On scales below the scalar sound-horizon, this approximation breaks down and the dark energy clusters together with dark matter \cite{Creminelli:2009mu,Sefusatti:2011cm}.

Each new operator  at order higher than quadratic introduces a new time-dependent function in the EFT action. Fortunately, under the assumptions outlined above one needs to introduce only three new operators and, correspondingly, time-dependent functions.
This small number of new functions is due to the second-order character of  Horndeski theories, combined with the requirement that the equations of motion are dominated by terms with the highest number of spatial derivatives.

To derive the three equations describing the relation among the two gravitational potentials, the dark-energy field,  and the matter fluctuations, we rewrite the nonlinear EFT action in Newtonian gauge by explicitly re-introducing the field fluctuations by a time-coordinate change. { The  coordinate change  required to expand the action in Newtonian gauge has to be performed at up to second order and we give the details  on how to do that  in App.~\ref{Stuek}.
We derive the Newtonian gauge action in the quasi-static regime in}
Sec.~\ref{NMG}, first in the linear and then in the nonlinear case. Using this action, we derive the effective Poisson equation necessary to solve the dark matter fluid dynamics in standard perturbation theory \cite{Bernardeau:2001qr}   
(see e.g.~\cite{Kimura:2011dc,Takushima:2013foa,Takushima:2015iha,Bellini:2015wfa}  for other approaches that include  modifications of gravity from Horndeski theories in standard perturbation theory). { This modified Poisson equation is used to compute the one-loop matter power spectrum within the Effective Field Theory of Large-Scale Structure (EFTofLSS) approach in~\cite{CLV2}. 

In Sec.~\ref{strongcouplingsec} we  discuss the different scales relevant for our study: the nonlinear scale of perturbations, the Vainshtein scale relevant for screening, and the strong coupling energy scale of dark energy fluctuations. In particular, we  show that the latter is typically much higher  than the dark matter nonlinear scale, which is consistent with the use of the EFTofDE  in the nonlinear regime of cosmological perturbations.  Finally, in Sec.~\ref{sphericalvainsec} we briefly discuss the Vainshtein scale around spherical sources and  compute when screening effects become important in the mildly nonlinear scales studied in \cite{CLV2}. 
For  earlier  studies of the Vainshtein screening in Horndeski theories expanded around a flat and cosmological background, see respectively \cite{Koyama:2013paa} and \cite{Kimura:2011dc,Kase:2013uja}. In Sec.~\ref{gw17} we comment on the consequences on our nonlinear operators of the recent simultaneous observation of gravitational waves and gamma ray bursts from the binary pulsars inspiral GW170817 \cite{TheLIGOScientific:2017qsa,Goldstein:2017mmi}. Finally, we conclude in Sec.~\ref{concref}.}

\section{The action} \label{eftofdesec}

When approaching short scales, gravitational as well as scalar field nonlinearities become  important. To study this regime, we must
extend the EFTofDE approach developed for linear perturbations in \cite{Gubitosi:2012hu,Gleyzes:2013ooa,Gleyzes:2014rba}  and
introduce higher-order operators (see also \cite{Frusciante:2017nfr}).

\subsection{Linear building blocks of Horndeski theories} \label{blocks}

Let us start reviewing the construction of the  EFTofDE quadratic action for Horndeski theories.
The dynamics of  Horndeski theories 
 is governed by the action 
\be
S_{\rm H}  = \int d^4x \sqrt{-g} {\mathcal{L}}_{\rm H}   \; ,
 \label{Horndeski}
\ee
with Lagrangian density \cite{Kobayashi:2011nu}
%
%
%
\be
\begin{split}
 \label{HorndeskiL}
 {\mathcal{L}}_{\rm H} &= G_2(\phi,X) +  G_3(\phi, X) \Box \phi   + G_4(\phi,X) \, {}^{(4)}\!R - 2 G_4{}_{,X}(\phi,X) (\Box \phi^2 - \phi^{ \mu \nu} \phi_{ \mu \nu})  \\
&+  G_5(\phi,X) \, {}^{(4)}\!G_{\mu \nu} \phi^{\mu \nu}  +  \frac13  G_5{}_{,X} (\phi,X) (\Box \phi^3 - 3 \, \Box \phi \, \phi_{\mu \nu}\phi^{\mu \nu} + 2 \, \phi_{\mu \nu}  \phi^{\mu \sigma} \phi^{\nu}_{\  \sigma})    \; ,
\end{split}
\ee
where we have used the following definitions: $\phi_\mu \equiv \nabla_\mu \phi$, $\phi_{\mu \nu} \equiv \nabla_\nu \nabla_\mu \phi$, $\Box \phi \equiv \nabla^\mu \nabla_\mu \phi$, and $X \equiv g^{\mu \nu} \nabla_\mu \phi \nabla_\nu \phi = \phi_\mu \phi^\mu  $.

In the following, we assume that the scalar field is always spacelike. When this is the case, as shown in \cite{Gleyzes:2013ooa} this action can be rewritten in  the language of the EFTofDE approach, in the so-called unitary gauge, where the time coordinate coincides with the hypersurfaces of uniform scalar field. In this gauge, the operators in the action transform as scalars under time-dependent spatial diffeomorphisms on the hypersurfaces of uniform scalar field. 

To derive this action, one can proceed in two steps, reviewed in App.~\ref{app:Horndeski}. The detailed derivation can be found in \cite{Gleyzes:2013ooa}. 
First, one can rewrite the action above in terms of geometrical quantities defined on the uniform scalar field hypersurfaces. A single derivative on $\phi$ can be rewritten in terms of the unit vector to the uniform-field hypersurfaces, $n_\mu \equiv \partial_\mu \phi/\sqrt{-X}$. Two derivatives on $\phi$ can be written as a covariant derivative on $n_\mu$, which we can   project on the spatial hyperfurfaces. This gives the extrinsic curvature $K^{\mu}{}_{ \nu} \equiv  h^{ \mu \rho} \nabla_\rho n_{\nu}$, where $h_{\mu \nu} \equiv g_{\mu \nu} + n_\mu n_\nu$ is the metric of the spatial hypersurfaces. Finally, we can use the Gauss-Codazzi relation to decompose the 4-dimensional curvature quantities (the Ricci scalar ${}^{(4)} R$, and Einstein tensor ${}^{(4)} G_{\mu \nu}$) in terms of the extrinsic curvature and the 3-dimensional Ricci tensor of the spatial hypersurfaces, that we denote by $R_{\mu \nu}$.  
After this long but straighforward decomposition, we can choose the unitary gauge. With this choice,  $X=g^{00}\dot { \phi}^2(t)$  and the action can be written as
\be
S_{\rm H}= \int d^4 x \sqrt{-g} {\mathcal{L}}_{\rm H} \left( t, g^{00}, K^{\nu}{}_\mu, R^{\nu}{}_\mu \right) \;.
\ee

In a second step, this action can be then expanded around a flat FLRW background metric, $ds^2 = -dt^2 + a^2(t) d\vec x^2$. In particular, we can define 3-dimensional tensors vanishing on the background: $\delta g^{00} \equiv 1+ g^{00}$ and $\delta K_{\mu \nu} \equiv K_{\mu \nu} - H h_{\mu \nu}$, where $H\equiv \dot a/a$ is the Hubble rate. Since the homogeneous universe has no spatial curvature, the Ricci tensor $R_{\mu \nu}$ is already a perturbed quantity.
Expanded at {\em quadratic order} in these quantities, the unitary gauge Horndeski action 
can be written  as \cite{Gleyzes:2013ooa,Gleyzes:2014rba}  
\be
\begin{split}
\label{quad_actionEFT}
& \int  d^4 x \sqrt{-g}  \bigg\{  \frac{\MM^2 f (t)}{2}  {}^{(4)}\!R  - \Lambda (t) - c (t) g^{00}    \\
 & 
 + \frac{m_2^4(t) }{2} (\delta g^{00})^2- \frac{m_3^3(t)}{2} \, \delta K \delta g^{00} + m_4^2 (t) \left( \d K^{\nu}{}_\mu \d K^{\mu}{}_{\nu}  - \delta K^2+  \frac{1}{2} \delta g^{00} \R \right)    \bigg\} \;,
\end{split}
\ee
where $\Lambda$, $c$, $m_2^4$, $m_3^3$ and $m_4^2$ are time dependent functions (note that  $m_2^4$, $m_3^3$ and $m_4^2$ are written as mass to some power to keep track of the dimensions but can have either signs).  For the matter sector, we describe dark-matter particles by a non-relativistic fluid minimally coupled to the gravitational metric $g_{\mu \nu}$ (the action is explicitly given below in \eqn{coupl_matt}). 
The background equations obtained by varying the first line of \eqn{quad_actionEFT} along with the background part of matter action read \cite{Gubitosi:2012hu}
\begin{align}
\label{back1}
c +\Lambda  & =3 \MM^2 \left(H  \dot f +f  H^2 \right)- \bar \rho _{\rm m} \;, \\
\label{back2}
\Lambda  -c & =\MM^2 \left( \ddot f  +2 H \dot  f +2 f \dot H +3 f  H^2\right) \;,
\end{align}
which show that $\Lambda$ and $c$ can be expressed in terms of $H$, $f$ and of $\bar \rho_{\rm m}$, the homogeneous matter energy density. Therefore,
the action is described in terms of {\em four} independent operators \cite{Gleyzes:2013ooa}.

Following the notation introduced in \cite{Bellini:2014fua}, 
we define the time-dependent effective Planck mass 
as
\be
M^2   \equiv \MM^2  f + 2 m_4^2 \;.
\ee
This sets the normalization of the tensor perturbations of the metric and has to be strictly positive.
Moreover, it is convenient to define  time-dependent functions, connected to the EFT parameters by
\cite{Gleyzes:2014rba}
\be
\label{EFTaction_masses}
 \alphaK \equiv \frac{2c + 4 m_2^4}{M^2 H^2}  \;, \qquad \alphaB \equiv \frac{\MM^2 \dot f - m_3^3 }{2 M^2 H} \;,  \qquad
 \alphaM  \equiv  \frac{ \MM^2 \dot f + 2 (m_4^2)^{\hbox{$\cdot$}}}{M^2 H }\;, \qquad  \alphaT  \equiv - \frac{2 m_4^2}{M^2 }  \;.
\ee
These quantities are related to the background values of the functions $G_I(\phi, X)$ of the Horndeski action \eqref{Horndeski}. The explicit relations are reported in  App.~\ref{app:Horndeski}.

When expanded at quadratic order in the metric perturbations, the above  action governs the linear dynamics of scalar and tensor fluctuations. The details on the derivation of their quadratic action can be found for instance in \cite{Gleyzes:2013ooa,Gleyzes:2014rba}.
The  dispersion relations of these propagating modes is respectively given by $\omega^2 = c_s^2 k^2$ for scalars and $\omega^2 = c_{T}^2 k^2$  for tensors. The speed of scalar fluctuations is given by
\be
\label{tildecss}
c_s^2 =  - \frac{2}{\alpha} \bigg\{  (1+\alphaB) \bigg[ \alphaB (1+\alphaT) + \alphaT  - \alphaM + \frac{\dot H}{H^2}   \bigg] + \frac{ \dot \alpha_{\rm B} }{H}  
+  \frac{ \bar \rho_{\rm m}}{2 M^2 H^2 }  \bigg\} \;, \qquad \alpha \equiv \alphaK + 6 \alphaB^2 \;.
\ee
The one for tensors is $c^2_{T} = 1 + \alphaT$. To avoid gradient instabilities, we require the speeds of propagation to be always positive. Moreover, the kinetic energy of the scalar mode is proportional to the parameter $\alpha$ defined in eq.~\eqref{tildecss}, which has to be positive  to avoid ghost instabilities \cite{Gleyzes:2013ooa}. In summary, we require the usual four stability conditions on the above parameters, i.e., 
\be
\label{stab_cond}
M^2 > 0 \;, \qquad \alphaT> -1 \;, \qquad \alpha>0 \;, \qquad c_s^2 >0 \;.
\ee


%
%
\subsection{Quasi-static limit}  

As explained in the introduction, we concentrate on scales much shorter than the Hubble radius, where relativistic effects due to  the expansion of the universe can be neglected. Moreover, we consider non-relativistic gravitational fields and velocities. For gravitational and field fluctuations below the scalar field sound horizon, we can assume the quasi-static approximation and time  derivatives can be taken to be much smaller than  spatial derivatives.  

For concreteness, let us consider an explicit gauge for the metric. In Newtonian gauge, focusing only on scalar perturbations, the metric reads
\be
\label{metric_Newtonian}
ds^2 = - (1+2 \Phi) dt^2  + a^2(t) (1-2 \Psi)  \delta_{ij} d x^i d x^j \;.
\ee
We then assume that dark energy is described by a dimensionless scalar field $\chi$ (this is explained in  more detail in Section~\ref{NMG}). 
{For convenience, here and below  we will use the following notation, 
\be
\varphi_1 \equiv \Phi \;, \qquad \varphi_2 \equiv \Psi \;, \qquad \varphi_3 \equiv \chi \; .
\ee
At leading order in the quasi-static expansion, considering theories where Vainshtein screening is the dominant screening mechanism, the relevant operators in the action are those  with the highest number of spatial derivatives per number of fields. 
More specifically, for  operators with $n$  fields, the dominant ones  contain $2(n-1)$ spatial derivatives.}
Fortunately, their number is finite. Because we restrict to  Horndeski theories, which have second order equations of motion both for the metric and the scalar field \cite{Horndeski:1974wa}, we expect the operators in the gravitational action to have the double $\varepsilon$ structure of the Galileons \cite{Nicolis:2008in}. In particular, they have either one of these forms (see also explicit expressions in~\cite{Kimura:2011dc,Kobayashi:2014ida}),
\be
\label{dominantQS}
\begin{split} 
 &\varepsilon^{ikl} \varepsilon^{j k l} \partial_i  \varphi_a \partial_j \varphi_b  \;, \\
 & \varepsilon^{ikm} \varepsilon^{jl m} \partial_i  \varphi_a \partial_j \varphi_b \partial_k \partial_l \varphi_c \;, \\
 & \varepsilon^{ikm} \varepsilon^{jl n} \partial_i \varphi_a \partial_j \varphi_b  \partial_k \partial_l \varphi_c \partial_m \partial_n \varphi_d   \;,
\end{split}
\ee
where $\varepsilon^{ijk}$ is the 3-dimensional Levi-Civita symbol.
Since in $3+1$-dimensions there are no operators with $2(n-1)$ spatial derivatives for $n> 4$, we can stop at quartic order in the number of fields. As we will see below, the number of independent  operators generating terms of the type \eqref{dominantQS} are only six.
{Other operators with less spatial derivatives, such as for instance
$
\varphi_a^2$, $\varphi_a (\partial \varphi_b)^2$ and $ (\partial \varphi_a)^2 (\partial \varphi_b)^2 $,
contribute to post-Newtonian corrections.}

The operators of the action \eqref{quad_actionEFT} can be expanded at higher order in the metric perturbations, so that they also contribute to the interaction Lagrangian.  However, in the quasi-static approximation, not all terms in this  expansion are relevant. We are going to see this in more detail in the next section, 
where we will move out of unitary gauge by reintroducing the scalar field fluctuations via the Stueckelberg trick and write  the action in Newtonian gauge. The explicit transformations  are given in App.~\ref{Stuek}.  Here we anticipate some of these results. 

As we will see, since the highest interactions that we need to consider are quartic, we will only need the expansion of 
$\delta g^{00}$, $\delta K^\mu{}_\nu$ and $R^\mu{}_\nu$ up to quadratic order.  Moreover, as mentioned above the relevant Lagrangian terms in the quasi-static limit have the form \eqref{dominantQS}.
In Newtonian gauge $\delta g^{00}$ does not introduce any spatial derivative at linear order, while at second  order it introduces two spatial derivatives. Schematically,
\be
\delta g^{00} \; \to \;  \varphi_a \;, \  ( \partial \varphi_a )^2    \;,
\ee
while for the extrinsic and intrinsic curvature tensors we have
\begin{align}
\delta K & \;  \to \;   \partial^2 \varphi_a \;,  \ ( \partial \varphi_a )^2     \;,  \\
 R & \;  \to \;  \partial^2 \varphi_a  \;,  \   ( \partial^2 \varphi_a )^2     \;.
\end{align}
Therefore,  in the quasi-static limit $\Lambda$, $c$ and $m_2^4$, and thus $\alphaK$,
do not contribute to the action at quadratic  order in perturbations. If $f$ is time independent, the operator $f {}^{(4)}\!R$ generates the usual quadratic action in the quasi-static limit and does not contribute at higher-order. When $f$ is time dependent, $f {}^{(4)}\!R$ contains the operator $ \dot f \delta g^{00} \delta K$, as can be shown upon use of the Gauss-Codazzi relation. This operator combines with
$m_3^3$  into  $\alphaB$ and generates terms of the form $(\partial \varphi _a)^2 \partial^2 \varphi_b$, which  contribute at most to the cubic order. 
Finally,  the term $\delta g^{00} R$ in the operator $m_4^2$ generates  relevant terms: $(\partial \varphi_a)^2 \partial^2 \varphi_b$ and $(\partial \varphi_a)^2 (\partial^2 \varphi_b)^2$. Thus, $\alphaT$ contributes both to the cubic and quartic interactions.

\subsection{Nonlinear building blocks}\label{nonlinear}

Going back to the EFT action, let us consider operators cubic and quartic in $\delta g^{00}$, $\delta K^\nu{}_\mu$ and $R^\nu{}_\mu$ and construct the full nonlinear action in the quasi-static limit. 
The detailed calculation is given in App.~\ref{app:Horndeski}, which also includes 
operators containing less spatial derivatives, which therefore do not contribute to the quasi-static limit. 

For convenience, let us define two quadratic combinations that systematically appear in the nonlinear expansion, 
\be\label{K2}
\begin{split}
\delta {\cal K}_2 & \equiv  \delta K^2 - \d K^\nu{}_\mu \d K^\mu{}_\nu  \;, \qquad
\delta {\cal G}_2  \equiv \delta K^\nu{}_\mu \R^\mu{}_\nu - \frac12 \d K \R \;, \\
\end{split}
\ee
and the cubic combination
\be\label{K3}
\delta {\cal K}_3  \equiv \delta K^3 -3\, \d K \d K^\nu{}_\mu \d K^\mu{}_\nu+ 2 \, \d K^\nu{}_\mu \d K^\mu{}_\rho \d K^\rho{}_\nu \;.
\ee
At cubic order, the expanded action contains five  operators. These are
\be
\begin{split}
\label{cubic_ops}
&(\delta g^{00})^3 \;, \qquad (\delta g^{00})^2 \delta K \; , \qquad (\delta g^{00})^2 \R \;, \qquad  \delta g^{00}  \delta {\cal K}_2  \;, \qquad  \delta {\cal K}_3   + 3  \delta g^{00} \delta {\cal G}_2 \;,
\end{split}
\ee 
 where only four are independent, because the coefficient in front of the third operator is linearly related to other coefficients which have already been defined. See derivation in App.~\ref{app:Horndeski}.
However, following the discussion above it is straightforward to verify that the first three operators do not contribute to the quasi-static limit.

At quartic order, instead, one has six operators,
\be
\begin{split}
\label{quartic_ops}
&(\delta g^{00})^4 \;, \qquad (\delta g^{00})^3 \delta K \; , \qquad (\delta g^{00})^3 \R \;, \qquad  (\delta g^{00})^2  \delta {\cal K}_2\;,  \qquad  (\delta g^{00})^2 \delta {\cal G}_2 \;, \qquad  \delta g^{00} \delta {\cal K}_3  \;.
\end{split}
\ee 
Only four of them are independent: the explicit expressions can be found in App.~\ref{app:Horndeski},
but only the last one contributes to the quasi-static limit.
At fifth order the  operators are still six and  can be obtained by multiplying the quartic operators by $\delta g^{00}$. The same logic applies to sixth and higher-order operators. However, none of them contribute to the quasi-static limit. 

In conclusion, the full nonlinear gravitational action for Horndeski theories in the quasi-static limit can be written in terms of only six independent operators, as
\be
\begin{split}
\label{total_actionEFT}
S_{\rm g}  = & \int  d^4 x \sqrt{-g}  \bigg\{ \frac{\MM^2 f (t)}{2}  {}^{(4)}\!R  - \frac{m_3^3(t)}{2} \, \delta K \delta g^{00}  
- m_4^2 (t) \left( \delta {\cal K}_2- \frac{ 1}{2} \,\delta g^{00} \R \right)    \\
 &    \hspace{1in} - \frac{m_5^2(t)}{2}  \delta g^{00} \delta {\cal K}_2-   \frac{m_6(t)}{3} (\delta {\cal K}_3   + 3  \delta g^{00} \delta {\cal G}_2)    -   \frac{m_7(t)}{3} \delta g^{00} \delta {\cal K}_3 \bigg\} \;,
\end{split}
\ee
where $m_5^2$, $m_6$ and $m_7$ are new time-dependent functions.
As for the linear theory, we can define  dimensionless time-dependent functions parametrizing the nonlinear action,
\be
\begin{split}
\label{EFTaction_masses2}
  \bone  \equiv \frac{2 m_5^2+ 2 H m_6}{M^2} \;, \qquad  \btwo \equiv \frac{2 H m_6}{M^2} \;, \qquad  \bthree \equiv \frac{4 H  m_7 +2 H m_6}{M^2}  \;.
\end{split}
\ee
The reason for the above combinations will become clear in Sec.~\ref{NMG}.

In terms of the functions $G_I(\phi,X)$ of the Lagrangian \eqref{HorndeskiL}, these are given by
\be
\begin{split}
\bone & = \frac{2 X}{M^2}  \Big[ 2G_{4,X} + 4 X G_{4,XX} +  2H \big(  \dot \phi G_{5,X} +   \dot \phi X G_{5,XX} \big)+  G_{5,\phi} +  X G_{5,X \phi}   \Big] \;,\\
\btwo & = - \frac{2 H }{M^2} \dot \phi X  G_{5,X}  \;, \\
 \bthree & = \frac{4 H}{M^2} \dot \phi X \Big( G_{5,X}+ X G_{5,XX}  \Big) \;. \label{alphas457}\end{split}
\ee
Quantities analogous to $\alphaq$ and $\alphac$ have been introduced by Bellini {\em et al.}~\cite{Bellini:2015wfa} and Yamauchi {\em et al.}~\cite{Yamauchi:2017ibz} in the calculation of the bispectrum. The relation between our notation and theirs is: $\bone = - 2 \alpha^{\rm (B)}_4 + 2 \alpha^{\rm (B)}_5$ and $\btwo = - \alpha^{\rm (B)}_5$ for the comparison with \cite{Bellini:2015wfa} and
$\bone = - 2 \alphaq^{\rm (Y)} - 2 \alphac^{\rm (Y)}$(\footnote{There is a typo in the expression of $\alphaq^{\rm (Y)}$ in eq.~(47) of Ref.~\cite{Yamauchi:2017ibz}. In their expression $G_{4,\phi}$ should be read $G_{5,\phi}$. We thank the authors of this reference for having checked and privately agreed on this correction. }) and $\btwo = - \alphac^{\rm (Y)}$ for the comparison with \cite{Yamauchi:2017ibz}.

\subsection{Matter coupling}\label{mattersec}

We now need to consider the coupling of the gravitational metric to dark matter. Assuming minimal coupling to $g_{\mu \nu}$, this is described by 
\be
S_{\rm m} = - \frac12  \int d^4 x \sqrt{-g}  T^{(\rm m)}_{\mu \nu} \, \delta g^{\mu \nu} \;,
\ee 
where $T^{(\rm m)}_{\mu \nu}$ is the stress-energy tensor of cold dark matter particles. 
In this work we
 describe cold dark matter in the perfect fluid approximation with vanishing pressure. The stress-energy tensor then reads
\be \label{matterstresstensor}
T^{(\rm m)}{}^{ \mu}{}_\nu = \rho_{\rm m} u^\mu u_\nu \;, 
\ee
where $ \rho_{\rm m}$ is the energy density in the rest frame of the fluid and $u^\mu$ its 4-velocity. At leading order in small velocities and fields we have  
$u^\mu =  (1/\sqrt{-g_{00}}, a^{-1} v ^i)$
and
\be
T^{(\rm m)}{}^0_{\ 0}  =  - \rho_{\rm m} \equiv  - \bar \rho_{\rm m} ( 1 + \deltam ) \;, \qquad T^{(\rm m)}{}^0_{\ i}  =  \rho_{\rm m} a   v ^i = - a^2 T^{(\rm m)}{}^i_{\ 0}\;, \qquad
T^{(\rm m)}{}^i_{\ j}  = \rho_{\rm m} v^i v^j\label{se}\;,
\ee
where we have defined $\delta $ and $ v^i$, respectively  the  energy density contrast and 3-velocity of matter. 
Using the metric in Newtonian gauge \eqn{metric_Newtonian} and dropping negligible terms in the short-scale limit, the matter coupling to gravity is described by 
\be
\label{coupl_matt}
S_{\rm m} = - \int d^3 x dt a^3 \bar \rho_{\rm m}  \, (1+\deltam) \,  \Phi \;.
\ee 
We stress that even though we are assuming non-relativistic perturbations, this expression holds also for large $\delta$: it is thus the fully nonlinear matter coupling.

Matter density and velocity satisfy the usual continuity and Euler equations. In order to solve them, we need a relation between $\Phi$ and $\delta$, which in general relativity   is given by the Poisson equation. As we will see in Sec.~\ref{nlt}, this relationship is altered and becomes nonlinear in the presence of dark energy.  Finding the expression for $\partial_i \Phi$ in terms of $\delta$ will be one of the main results of this paper.

\subsection{Frame dependence of the action} \label{sec:frame}

One can describe the gravitational action $S_{\rm g}$  by using a different metric. For instance, one can consider the metric redefinition
\be
g_{\mu \nu} \  \to \  \hat g_{\mu \nu} = C(\phi) g_{\mu \nu} + D(\phi) \partial_\mu \phi \partial_\nu \phi\; , \label{disftran}
\ee
which is a  disformal transformation with conformal and disformal factors that depend on $\phi$, and rewrite the action in terms of  
$\hat g_{\mu \nu}$. Of course, the matter coupling to the new gravitational metric $ \hat g_{\mu \nu}$ is different from the coupling to $g_{\mu \nu}$.
For instance, if $g_{\mu \nu}$ is the Jordan frame metric, where matter is minimally coupled, in the frame of $\hat g_{\mu \nu}$ matter is explicitly coupled to the scalar.

It has been shown that the structure of the Horndeski action is preserved under the transformation above \cite{Bettoni:2013diz}. Therefore, we expect that the structure of $S_{\rm g}$ is as well preserved under this transformation. Indeed,   the transformed action has the same operators of \eqref{quad_actionEFT}, but with different coefficients. Their transformations have been given explicitly in \cite{Gleyzes:2015pma} in terms of how  $M^2$, $H$ and the time-dependent functions $\alphaK$, $\alphaB$, $\alphaM$ and $\alphaT$ change. This reads
\be
\label{frame_tra1}
\begin{split}
\hat M^2 & = \frac{M^2}{C(1+\alphaD)} \;, \qquad \hat H =  H (1+\alphaC) \sqrt{\frac{1+\alphaD}{C}} \;, \\ 
\hat \alpha_{\rm B} & = \frac{1 + \alphaB  }{(1+\alphaC) (1+\alphaD) } - 1  \;, \\
\hat \alpha_{\rm M} & =  \frac{\alphaM - 2 \alphaC}{1+\alphaC} -  \frac{\dot \alpha_{\rm D}}{ 2 H  (1+\alphaD) (1+\alphaC)}\;, \\
\hat \alpha_{\rm T} & =  (1+\alphaT) (1+\alphaD) - 1  \;.
\end{split}
\ee
Here $\alphaC$ and $\alphaD$ are  time-dependent functions respectively parametrizing the conformal and disformal transformation and defined by 
\be
\alphaC \equiv \frac{\dot \phi}{2 H } \frac{d \ln C}{d \ln \phi} \;, \qquad \alphaD \equiv - \frac{D}{D + C/X}\;, 
\ee
where the right-hand side is evaluated on the background.

We then focus on how the nonlinear part of the action \eqref{total_actionEFT} transforms.  In particular, the time-dependent functions $\alphaq$, $\alphac$ and $\alphaset$ are redefined by the transformation \eqref{disftran} as
\be
\label{frame_tra}
\begin{split}
\hatbone & = \frac{\bone   + \alphaD+ 2 \btwo ( \alphaC + \alphaD + \alphaC \alphaD)}{1+\alphaD }    \;, \\
\hatbtwo & = \btwo (1+\alphaC) \;, \\
\hatbthree & = (1+\alphaC)  \frac{\bthree + 3 \alphaD \btwo}{1+\alphaD }   \;.
\end{split}
\ee
We notice that, starting from a theory without nonlinear couplings, $\alphaq=\alphac=\alphaset=0$, we can generate only $\hat {\alpha}_{\rm V1}$ by the transformation \eqref{disftran}.

\section{Nonlinear equations} \label{NMG}

Focusing on the operators that dominate the short-scale regime, we now derive 
the nonlinear  action and the relevant equations in the Newtonian gauge, assuming minimal coupling of matter. 
As a warm-up and to set up the notation, let us start discussing the quadratic action and the linear equations. 

\subsection{Linear theory}

We first want to move from unitary gauge and to do so we introduce the scalar field fluctuation $\pi$ via the time diffeomorphism $t \to t+ \pi (t,\vec x)$. For convenience, instead of $\pi$ we will use the dimensionless field $\chi$, defined as 
\be
\label{defchi}
\chi \equiv H \pi \;.
\ee
We  assume a perturbed FLRW metric in Newtonian gauge, see eq.~\eqref{metric_Newtonian}.
The effect of a time coordinate transformation on $g^{00}$, $\delta K^\nu{}_\mu$ and $R$ is derived in App.~\ref{Stuek}, up to second-order in the perturbations. Here we need only the linear  transformations, i.e.,
\be
\begin{split}
\label{variationRlin}
\delta K \to & \ \delta K  - 3  \dot H \pi  -   a^{-2} \partial^2 \pi    +{\cal O} (\pi^2) \;, \\ 
\delta K_{\ j}^i \to & \ \delta K_{\ j}^i - \dot H \pi  \delta^{i}_{j} -  a^{-2}   \partial_i \partial_j \pi + {\cal O} ( \pi^2) \;, \\
R \to & \ R + 4 a^{-2} H \partial^2 \pi     + {\cal O} (\pi^2) \;.
\end{split}
\ee

Following the notation introduced in Sec.~\ref{eftofdesec}, in order to write the quadratic action in compact form we use the vector $\varphi_a$ ($a=1,2,3$),
\be \label{phiadef}
\varphi_a\equiv\left(
\begin{array}{c}
\Phi\\
\Psi\\
\chi
\end{array}
\right)\,.
\ee
Using eq.~\eqref{variationRlin} to transform the  unitary gauge action  \eqref{quad_actionEFT} and keeping only terms quadratic in the perturbations, this becomes
\be
\begin{split}
\label{total_action_pert_quad}
S_{\rm g}^{(2)}   = - \int d^3 x dt \,   a \, M^2  A_{ab} \partial_i \varphi_a \partial_i \varphi_b  \;,
\end{split}
\ee
where $A_{ab}$ is a dimensionless symmetric matrix of components
\be
A_{ab}=\left(
\begin{array}{ccc}
0&1&-\alphaB\\
1&-1-\alphaT&\alphaM-\alphaT\\
-\alphaB& \alphaM-\alphaT&-\mathcal{C}_2
\end{array}
\right)\,,
\ee
and we have used  the functions $\alpha_I$, defined  in terms of the EFT parameters in eq.~\eqref{EFTaction_masses}.
Moreover,  ${\cal C}_2$    is a  time-dependent function  explicitly given by
\be
\label{C2}
\begin{split}
{\cal C}_2  &=   \alphaT - \alphaM  + \alphaB  ( 1 +\alphaM) + (1+\alphaB ) \frac{ \dot H }{ H^2} +  \frac{\dot \alpha_{\rm B} }{H} +  \frac{\bar \rho_{\rm m}}{2 H^2 M^2}  \;,
\end{split}
\ee
which is obtained after use of the background equations \eqref{back1} and \eqref{back2} to replace  $c(t)$ in the action.

In terms of the vector $\varphi_a$,
 the quadratic coupling to matter \eqref{coupl_matt} reads
\be
S_{\rm m} =  - \int d^3 x dt  a^3 \bar \rho_{\rm m}  \varphi_1 \deltam \;.
\ee 
The variation of the sum of  $S^{(2)}+S_{\rm m}$ then gives the linear equations of motion
\be\label{masterapp_1}
0=\frac{1}{a2 M^2} \frac{\delta (S^{(2)} + S_{\rm m})}{\delta \varphi_a} = A_{ab}\partial^2 \varphi_b-  \deltac_{1 a} \, \frac{\bar \rho_{\rm m} a^2 }{2 M^2} \deltam \, ,
\ee
where $\deltac_{ab}$ denotes a Kronecker delta in field space.
In particular, the variation with respect to $\varphi_1= \Phi$, $\varphi_2 = \Psi$ and $\varphi_3=\chi$ are respectively equivalent to the quasi-static limit of the $(00)$ and $(ij)$ components of the full Einstein equations and Klein-Gordon equation for the scalar field.

The solution of eq.~\eqref{masterapp_1} for the Laplacian of the fields is 
\be
\label{linear_sol}
\partial^2 \varphi_a= \frac{\bar \rho_{\rm m} a^2 }{2 M^2} \Inv_{a 1} \deltam   \;,
\ee
where $\Inv_{ab}$ is the inverse matrix of $A_{ab}$. Its components can be written as
\be
\label{inv_A}
\Inv_{ab}  =\left(
\begin{array}{ccc}
1+\alphaT + \frac{ \xi^2}{\nu}&1+ \frac{ \xi \alphaB}{\nu}&  \frac{ \xi}{\nu}\\
1+ \frac{\xi \alphaB}{\nu}&\frac{\alphaB^2}{\nu}& \frac{  \alphaB}{\nu}\\
 \frac{ \xi}{\nu} & \frac{  \alphaB}{\nu} &\frac{1}{\nu}
\end{array}
\right)\,,
\ee
where we have introduced two dimensionless time-dependent functions, $\xi$ and $\nu$, given as
\be
\label{nuxi}
\xi = \alphaB (1+\alphaT) + \alphaT - \alphaM\;, \qquad  \nu = - {\cal C}_2 - \alphaB \left( \xi +   \alphaT - \alphaM \right)   \;.
\ee
Using the definition of the sound speed, eq.~\eqref{tildecss} and eq.~\eqref{C2}, $\nu$ can be rewritten as 
\be
\label{nucsa}
 \nu =  \frac{c_s^2 \alpha}{2} >0 \;,
\ee
where for the last inequality we have used the last two stability conditions \eqref{stab_cond}.
Specifically, rewriting the solution \eqref{linear_sol} using the explicit components of the inverse matrix in eq.~\eqref{inv_A}, we find
\begin{align}
\partial^2 \Phi & = \frac{\bar \rho_{\rm m} a^2 }{2 M^2}  \,  \mu_{\Phi} \,  \deltam \;, \qquad \mu_{\Phi} \equiv  \Inv_{11} =   1+\alphaT + \frac{ \xi^2}{\nu}  \;, \label{muPhi1}\\
 \partial^2 \Psi & = \frac{\bar \rho_{\rm m} a^2 }{2 M^2} \, \mu_{\Psi}\,  \deltam  \;, \qquad \mu_{\Psi} \equiv   \Inv_{12} =   1+ \frac{ \xi \alphaB}{\nu}  \;, \label{muPsi1}\\
 \partial^2 \chi & = \frac{\bar \rho_{\rm m} a^2 }{2 M^2} \, \mu_{\chi}\, \deltam  \;, \qquad \mu_{\chi} \equiv   \Inv_{13} = \, \frac{ \xi}{\nu}  \;, \label{muchi1}
\end{align}
where we have defined the time-dependent functions $\mu_{\Phi}$, $\mu_{\Psi}$ and $\mu_{\chi}$. In the case of standard gravity $\mu_{\Phi}=1$, $\mu_{\Psi}=1$ and $\mu_{\chi}=0$.
A generalization of these expressions to different species non-minimally coupled to the metric and to the case of beyond-Horndeski theories can be found, respectively, in Refs.~\cite{Gleyzes:2015pma} and \cite{DAmico:2016ntq}.

\subsection{Nonlinear theory} \label{nlt}

We can now expand the nonlinear action in the quasi-static approximation, eq.~\eqref{total_actionEFT}, to cubic and quartic order in the perturbations. 
{We obtain
\begin{align}
\label{total_action_pert_c}
S_{\rm g}^{(3)}  &= \int d^3 x dt \,  \frac{M^2}{ 3! \, a H^2} B_{abc} \, \varepsilon^{ikm} \varepsilon^{jl m} \partial_i  \varphi_a \partial_j \varphi_b \partial_k \partial_l \varphi_c  \;,  \\
S_{\rm g}^{(4)} &=  \int d^3 x dt \, \frac{M^2}{ 4! \,a^3 H^4} C_{abcd}\,   \varepsilon^{ikm} \varepsilon^{jl n} \partial_i \varphi_a \partial_j \varphi_b  \partial_k \partial_l \varphi_c \partial_m \partial_n \varphi_d \label{total_action_pert_q}\;,
\end{align}
where $\varepsilon^{ijk}$ is the 3-dimensional Levi-Civita symbol and} 
$B_{abc}$  and  $C_{abcd}$ are dimensionless time-dependent arrays, parametrizing the coupling strength between fields. They are symmetric under exchange of the arguments and their non-vanishing  elements     are
\be
\begin{split} \label{nlcoeffs}
B_{123} &=  B_{312} =B_{231}=B_{213}=B_{321}=B_{132}={\btwo}\,,\\ 
B_{133} & = B_{313}=B_{331}= {\bone} \,, \\
B_{233} & = B_{323}=B_{332} =  {\cal C}_3     \,, \qquad B_{333}  ={\mathcal{C}_4}  \,, \\
 C_{1333} &=  C_{3133}= C_{3313}=C_{3331} =- \bthree  \,,\qquad 
 C_{3333}  = \mathcal{C}_5  \,,
\end{split}
\ee
where we have introduced the following combinations of parameters,
\begin{align}
{\cal C}_3  & \equiv - \alphaT  - \btwo (1-\alphaM) - \btwo \frac{\dot H}{H^2} + \frac{ \dot \alpha_{\rm V2}}{H}    \;, \\
{\mathcal{C}_4} &\equiv - 4 \alphaB+ 2 \alphaM - 3 \alphaT   -( \bone +   \btwo )(1-\alphaM) - 3 \btwo \frac{\dot H}{H^2}  +  \frac{ \dotbone +  \dotbtwo }{H} \;,  \label{C4}\\
\mathcal{C}_5 & \equiv  3\left(  \alphaT -  \bone+\btwo + \bthree \right) -  (3 \btwo + \bthree )\alphaM + (3 \btwo + \bthree) \frac{\dot H}{H^2} - \frac{3 \dotbtwo  + \dotbthree}{H} \;. 
\end{align}
The specific form of the  interactions in eqs.~\eqref{total_action_pert_c} and \eqref{total_action_pert_q} is dictated by the structure of  the Horndeski Lagrangian; in particular, as shown below they lead to equations with exactly two spatial derivatives per field.
As discussed above, in the quasi-static limit the action at order higher than four vanishes. Thus, the sum of $S^{(2)}$, $S^{(3)}$ and $S^{(4)}$ represents the full nonlinear action in the non-relativistic limit.

We can now derive the full nonlinear equations obtained by varying the total action $S = S_{\rm g}^{(2)}+ S_{\rm g}^{(3)} + S_{\rm g}^{(4)} +S_{\rm m}$.
These can be written in compact form as 
\begin{align}\label{masterapp}
0& =\frac{1}{a2 M^2} \frac{\delta S}{\delta \varphi_d} \\
& = A_{da}\partial^2 \varphi_a-  \deltac_{d1} \,\frac{\bar \rho_{\rm m} a^2   }{2 M^2} \, \deltam   - \frac{B_{dab}}{4 H^2 a^2} \varepsilon^{ikm} \varepsilon^{jl m}  \partial_i \partial_j \varphi_a \partial_k \partial_l \varphi_b-\frac{C_{dabc}}{ 12 H^4 a^4}  \varepsilon^{ikm} \varepsilon^{jl n} \partial_i \partial_j \varphi_a  \partial_k \partial_l \varphi_b \partial_m \partial_n \varphi_c \nonumber \,.
\end{align}
Once expressed in terms of the original Horndeski functions, these nonlinear equations are equivalent to those obtained in Ref.~\cite{Kimura:2011dc}. 

 As we mentioned in Sec.~\ref{mattersec}, a main goal of this paper is to find the nonlinear relationship between $\partial^2 \Phi$ and $\delta$ that is needed to solve continuity and Euler equations for matter.  Because those equations will be solved perturbatively, we can solve \eqn{masterapp}  for small $\delta$ as well. With an eye toward computing the one-loop power spectrum, we need its solution up to third order. Formally, this reads
\begin{align}
\label{sol_NL}
\partial^2 \varphi_a= & \ H^2 a^2 \bigg\{ \frac{3 \, \omegam}{2}  \,  \mu_{\varphi_a }  \deltam +  \left( \frac{3\, \omegam}{2}  \right)^2 \mu_{\varphi_a,2}  \left[\deltam^2-\left(\partial^{-2}{\partial_i\partial_j}\deltam\right)^2\right]  \\
&+\left(  \frac{3\, \omegam}{2} \right)^3 \mu_{\varphi_a,22}\left[\deltam-\left(\partial^{-2} {\partial_i\partial_j} \deltam\right) \partial^{-2} {\partial_i\partial_j} \right]\left[\deltam^2-\left(\partial^{-2}{\partial_k\partial_l} \deltam\right)^2\right] \nonumber \\
&+\left(  \frac{3\, \omegam}{2}  \right)^3 \mu_{\varphi_a,3}\left[\deltam^3-3\deltam\left(\partial^{-2}{\partial_i\partial_j} \deltam\right)^2+2 (\partial^{-2}{\partial_i\partial_j} \deltam ) (\partial^{-2}{\partial_k\partial_j}\deltam )( \partial^{-2}{\partial_i\partial_k}\deltam)\right] \bigg\}  +  {\cal O} (\delta^4) 
\nonumber \,,
\end{align} 
where we have defined the time dependent fractional matter energy density,
\be
\Omega_{\rm m} \equiv \frac{\bar \rho_{\rm m} }{3 M^2 H^2} \;.
\ee
Notice that this definition differs from the standard one, as we have normalized the energy density to the effective Planck mass $M^2$, which can depend on time.
The explicit expressions for $\mu_{\varphi_a}$ are given in eqs.~\eqref{muPhi1}, \eqref{muPsi1} and \eqref{muchi1}. For the higher order terms, $\mu_{\varphi_a,2}$, $\mu_{\varphi_a,22}$ and $\mu_{\varphi_a,3}$ are given by
\be
\begin{split}
\mu_{\varphi_a,2} &= { \frac{1}{4} }  \Inv_ {ad} \, \Inv_ {1 c} \, \Inv_{1b} \, B_{dcb}\,,\\
\mu_{\varphi_a,22}& = { \frac{1}{8} }   \Inv_{ad} \, \Inv_{1 c} \, \Inv_{1 f} \, \Inv_{1 g} \, \Inv_{be} \, B_{d c b}  \, B_{e f g}\,,\\ 
\mu_{\varphi_a,3} & = { \frac{1}{12} }  \Inv_{ad} \, \Inv_{1e} \, \Inv_{1b} \, \Inv_{1c} \, C_{debc}\,.
\end{split}
\ee
 Indeed, these expressions vanish for standard gravity, as expected.  For $\varphi_a = \Phi$, these are explicitly given by
 \begin{align}\label{muphiexp}
 \begin{split}
\mu_{\Phi,2} & = { \frac{\mu_{\chi }}{4} }   \Big( 6  \mu_{\Phi} \mu_{\Psi} \btwo + 3 \mu_\chi \mu_\Phi  \bone  + 3 \mu_\chi \mu_\Psi   {\cal C}_3  + \mu_\chi^2 {\cal C}_4\Big) \;, \\
{ \mu_{\Phi,22} } & =  \frac{1}{8}   \bigg\{ 5 \mu_\Phi \mu_\chi^2 (\mu_\chi \bone  + 2 \mu_\Psi \btwo )^2 + 2 \mu_\chi^3 (3 \mu_\Psi {\cal C}_3 + \mu_\chi {\cal C}_4) (\mu_\chi \bone + 2 \mu_\Psi \btwo) \\
& \quad+\frac{1}{\nu} \big[ 2 \btwo \muphi ( 2 \mupsi- 1) +     2 \bone \muchi \muphi  + (3 \mupsi-1)  {\cal C}_3 \muchi  +  {\cal C}_4     \muchi^2 \big]^2    \bigg\}    \;,       \\
\mu_{\Phi,3} & =  { \frac{\mu_\chi^3}{12} }    \big( - 4 \mu_\Phi \bthree + \mu_\chi {\cal C}_5 \big) \;.
\end{split}
\end{align}


%

\subsection{Time-dependent couplings } \label{muplots}

In this section, we give a sample of the size of the effects of the EFTofDE couplings on the parameters $\mu_{\Phi,2}$, $\mu_{\Phi,22}$, and $\mu_{\Phi,3}$, given in \eqn{muphiexp}, which are relevant for dark matter clustering.  Because there are many parameter combinations, we do not present a detailed study of all of the possibilities here, but we focus on the nonlinear couplings $\alphaB$, $\alpha_{\rm V1}$, $\alpha_{\rm V2}$, and $\alpha_{\rm V3}$, so that we set $\alphaM = 0$ and $\alphaT = 0$.  
The first thing to notice is that $\alpha_{\rm V3}$ only enters in the expression for $\mu_{\Phi,3}$, so it does not affect $\mu_{\Phi,2}$ or $\mu_{\Phi,22}$.  

In order to investigate these functions, we  choose to parametrize the time-dependence of  $\alpha_I$  and ${\dot H}/{H^2}$ by taking
\be
\alpha_I ( a ) = \alpha_{I,0} \frac{1 - \omegam (a) }{1 - \Omega_{\rm m,0}}\ ,  \qquad  \frac{\dot H}{H^2} (a)= -\frac32 \Omega_{\rm m} (a)  \;, \qquad \Omega_{\rm m} (a)= \frac{ \Omega_{\rm m,0}  }{  \Omega_{\rm m, 0}  + (1 - \Omega_{\rm m, 0}) (a/a_0)^3  } \;, 
\ee
where for the fractional energy density of dark matter today we set $\Omega_{\rm m,0}= 0.281$.  In this way, the dark energy has a vanishing effect in the past, which is reasonable because  CMB experiments place fairly tight constraints at early times, and the background evolution is that of $\Lambda$CDM.

\begin{figure}[htb!]
\includegraphics[width=8cm]{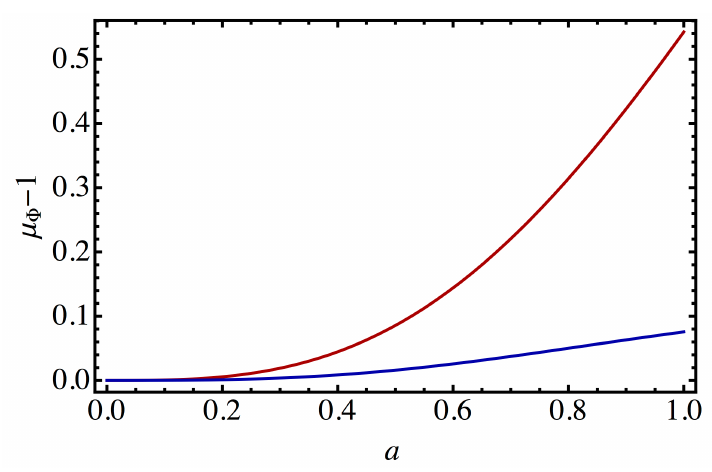}  \hspace{-0.07in} \includegraphics[width=8.2cm]{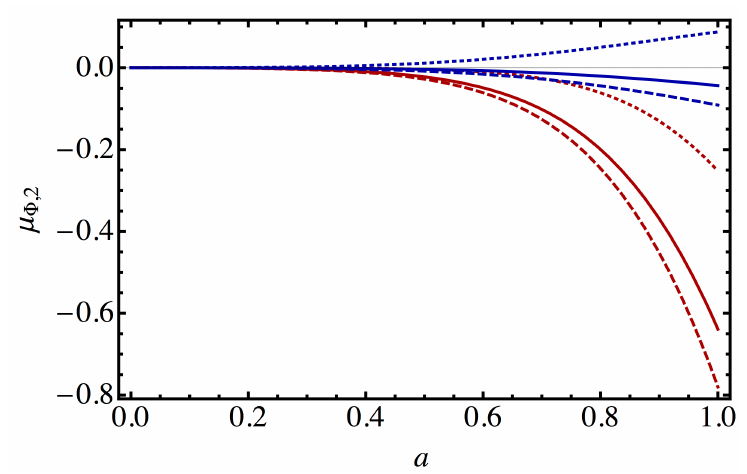} \\

\includegraphics[width=8cm]{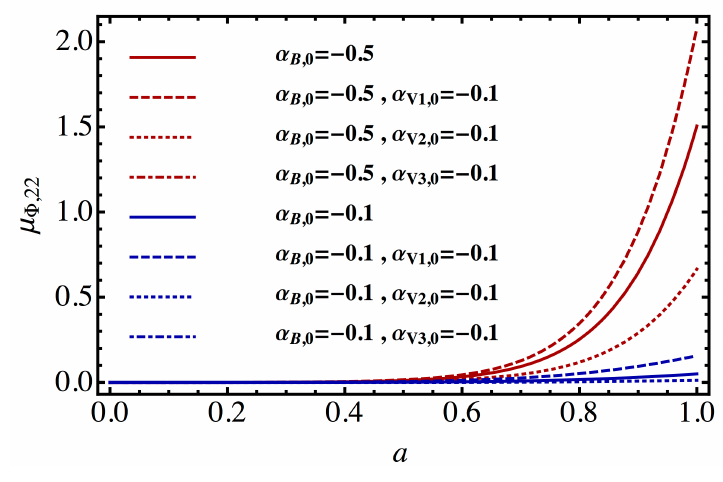}   \hspace{-.07in}     \includegraphics[width=8.2cm]{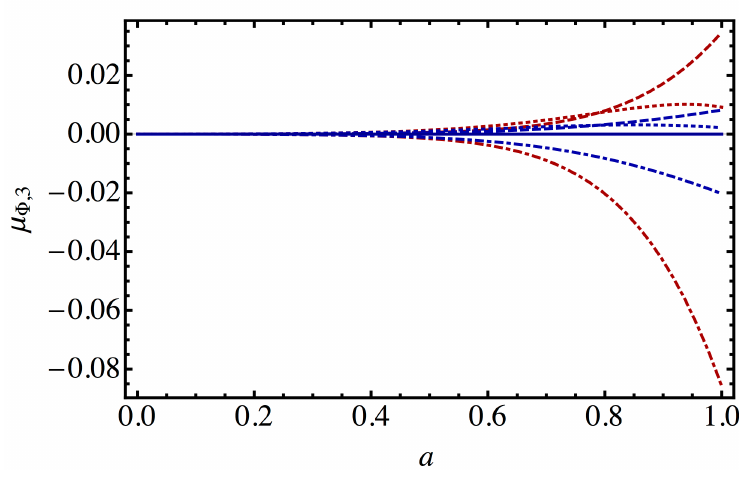}

\caption{  \footnotesize  Linear deviation $\mu_\Phi -1$ and nonlinear coupling functions $\mu_{\Phi,I}$, entering as coefficients to the nonlinear vertices in the dark-matter clustering equations, as a function of $a$ (see  \eqn{muphiexp}).  Note that $\alpha_{\rm V3}$ does not enter in $\mu_{\Phi,2}$ or $\mu_{\Phi,22}$ and that for 
$\alphaT =0 $, all contributions to  $\mu_{\Phi,3}$ are proportional to either $\alphaq$, $\alphac$ or $\alphaset$, which is why the solid red and  blue lines in the $\mu_{\Phi,3}$ plot are both at zero.  } \label{muplot}
\end{figure}

In Fig.~\ref{muplot}, we show a sample of the time dependent  functions $\mu_\Phi-1$ and $\mu_{\Phi,I} $ for different values of $\alphaB$, $\alpha_{\rm V1}$, $\alpha_{\rm V2}$, and $\alpha_{\rm V3}$.  We choose values of $\alpha_{\rm B,0}$ so that, for $\Omega_{\rm m,0} = 0.281$, the speed of sound of scalar propagation in \eqn{tildecss} is larger than zero.  Since the standard vertices appearing in the continuity and Euler equations are of order unity, the new vertices proportional to $\mu_{\Phi,I}$ will have a significant contribution when $\mu_{\Phi,I} \sim \mathcal{O}(1)$, although this depends on the computed statistics and 
its particular configuration in Fourier space.
From Fig.~\ref{muplot}, we see that in all three cases, the effect is larger when $|\alpha_{\rm B,0}|$ is larger, which  implies that there is also a significant change to the linear equations.  If one wants a small change to the linear equations and a large change to the nonlinear terms, this requires, for example, choosing $\alpha_{\rm B,0} = -0.1$ and $\alpha_{\rm \rm V3} = -4$, for which $\mu_{\Phi,3} ( a_0 ) \approx -1$.  We leave a detailed study of the various combinations of parameter  for future work.

%
%
%
%

\section{Nonlinear regime}  \label{NLreg}
\subsection{Strong-coupling scale, nonlinear scale, and Vainshtein scale}  \label{strongcouplingsec}

{Let us discuss the   scales that characterize our system.  
We first discuss the strong-coupling scale, i.e.~the energy scale at which nonlinear interactions of \eqref{total_actionEFT} exit perturbative unitarity. Above this scale, the EFTofDE is invalid and details about the UV completion, like the presence of additional degrees of freedom, become important. This scale depends on which of the nonlinear interactions is relevant in the particular setup. Since Horndeski theories have second-order equations of motion and reduce to the Galileons \cite{Nicolis:2008in} in the decoupling limit \cite{Koyama:2013paa}, the lowest $\LambdaU$  is obtained when high-derivative operators  are relevant.  In this case the strong-coupling scale is typically above $1000$ km \cite{Luty:2003vm}, roughly 18 orders of magnitude larger than the nonlinear scale of LSS. Another situation is when the nonlinear interactions are dominated by powers of $\delta g^{00}$, such as for the Ghost Condensate, in which case the strong-coupling scale is much  higher \cite{ArkaniHamed:2003uy}. In all cases we can comfortably assume that the EFTofDE is valid well above the nonlinear scale of structure formation.

Let us next discuss the nonlinear scale of dark matter clustering. In Ref.~\cite{CLV2} we use the EFTofLSS approach to compute the one-loop matter power spectrum.
In this context, there exists a scale, the nonlinear clustering scale $\knl$, at which the perturbative solution to the effective fluid equations breaks down \cite{Baumann:2010tm,Carrasco:2012cv}.  Indeed, the EFTofLSS is a controlled expansion in $k / \knl \ll 1$; near and above the scale $\knl$, the EFT does not converge and thus loses all predictability.  Below $\knl$, one can increase computational precision by including more and more loop corrections and can estimate the theoretical error by  the size of the next loop corrections which have not been included in the computation.  

Anticipating some of the results of Ref.~\cite{CLV2}, when computing the dark matter clustering in perturbation theory in the presence of modifications of gravity, the nonlinear effects can have two sources. First, there are the 
nonlinear vertices in the continuity and Euler equations. Since we assume minimal coupling of matter, these are the same as in $\Lambda$CDM and become important when density fluctuations become of order unity, i.e., 
\be
\label{deltaone}
\delta \sim 1 \;, \qquad k \sim \barknl  \;.
\ee
Secondly, there are the  nonlinear operators in the dark-energy action \eqref{total_actionEFT}, which lead to nonlinear terms in the generalized Poisson equation \eqref{sol_NL}. These nonlinearities are associated to the so-called Vainshtein screening mechanism \cite{Vainshtein:1972sx}, which we describe in more detail in the next subsection.
We denote by $\kV$, usually called the Vainshtein scale, the scale at which the nonlinearities of the scalar fluctuations become important, i.e.,
\be
\label{chione}
\frac{\partial^2 \chi}{H^2 a^2}  \sim 1  \;, \qquad k \sim \kV  \;.
\ee
Below this scale the system of equations \eqref{masterapp} is dominated by its nonlinear terms, and instead of the perturbative solution \eqref{sol_NL}, one finds that scalar fluctuations are suppressed, as we show below.

These two scales need not  be the same \cite{Fasiello:2017bot} and we distinguish two cases. If $ \barknl \ll \kV$, then screening is weak and does not take place until beyond the nonlinear scale for clustering. 
In this case the perturbative expansion is analogous to the one without modifications of gravity, but there can be modifications due to the different linear evolution for $\mu_\Phi \neq 1$.
On the contrary, if $ \kV \ll \barknl$, then the screening is strong and supplied by  nonlinear terms in the dark-energy action.  Thus, our computation is most useful in the intermediate regime where scalar-field nonlinearities are perturbative on mildly nonlinear scales.  Let us now look at the Vainshtein scale in more detail.}

\subsection{Spherically symmetric solutions} \label{sphericalvainsec}

We can study the Vainshtein screening regime \cite{Vainshtein:1972sx} by considering a situation where density perturbations are spherically distributed around the origin. In this case, all fields depend only on time and on the distance from the origin $r$. 
Let us define
\be
x(a,r) \equiv \frac{1}{a^2 H^2} \frac{\chi '}{r} \;, \qquad y(a,r) \equiv \frac{1}{a^2 H^2} \frac{\Phi '}{r} \;, \qquad z(a,r) \equiv \frac{1}{a^2 H^2} \frac{\Psi '}{r} \;, \qquad A(a,r) \equiv \frac{1}{M^2 H^2} \frac{m}{8 \pi r^3} \;,
\ee 
(here $\pi$ is the irrational number $3.14159$) with
\be
m(a,r) \equiv 4 \pi \int^r_0 \tilde r^2 \delta \rho_{\rm m} (a , \tilde r) d \tilde r \;.
\ee
By assuming spherical symmetry  and applying Stokes theorem,  eq.~\eqref{masterapp} can be rewritten as
\begin{align}
z  - \alphaB  x   - \btwo x z - \frac12 \bone  x^2   + \frac16 \bthree x^3 & = A \;,  \label{Vain1}\\
 (\alphaM - \alphaT) x +  y - (1+\alphaT) z     -\btwo x y     - \frac12 {\cal C}_3 x^2 & =0 \;, \label{Vain2} \\
  {\cal C}_2 x  + \alphaB   y  - ( \alphaM - \alphaT    ) z    + \frac12 {\cal C}_4 x^2  + \bone x y  + \btwo y z   
 +{\cal C}_3 x z     + \frac16 {\cal C}_5 x^3  -  \frac12 \bthree x^2 y & =0  \;. \label{Vain3}
\end{align}
For small density fluctuations, $A \ll 1$, we find the solutions (see eqs.~\eqref{muPhi1}, \eqref{muPsi1} and \eqref{muchi1})
\be
\label{lin_spher}
x  =  \mu_{\chi} A  + {\cal O}(A^2) \ , \hspace{.2in} y =  \mu_\Phi A + {\cal O}(A^2)  \ , \hspace{.2in} z =  \mu_\Psi A + {\cal O}(A^2)  \qquad (A\ll 1) \ .
\ee
Here we want to solve these equations for large $A$, $A \gg 1 $, where the field nonlinearities become important, triggering the Vainshtein regime.

Since eqs.~\eqref{Vain1} and \eqref{Vain2} are linear in $y$ and $z$, they can be solved for these two variables and the solution used in eq.~\eqref{Vain3}. In general, this yields a sextic equation for $x$. For $\alphac=0$, $\alphac=\alphaset=0$, and $\alpha_{\rm T} = \alphac=\alphaq=\alphaset=0$ this becomes,  respectively, a  quintic, a cubic,  and a quadratic equation.  For simplicity, we are going to discuss only the last two cases.

%
%
\subsubsection{Cubic screening}

 Let us consider   $\alpha_{\rm T} = \alphac=\alphaq=\alphaset=0$.
The quadratic equation for $x$ is
\be \label{quadraticeq}
 \frac{ {\cal C}_4}{2 \nu } x^2- x +   \frac{\xi}{\nu} A =0 \;,
\ee
where $ \xi$, $\nu$  and ${\cal C}_4$ are respectively defined in eqs.~\eqref{nuxi} and  \eqref{C4}. For our choice of parameters, $\xi$ and ${\cal C}_4$ simplify to $\xi = \alphaB -\alphaM$ and $ {\cal C}_4 = - 4 \alphaB + 2 \alphaM$.

 The  solution to this equation converging to the physical one  for  small $A$,  eq.~\eqref{lin_spher},  is
\be
x= \frac{\nu - \sqrt{\nu^2 - 2 A {\cal C}_4 \xi }}{{\cal C}_4} \;.
\ee
Assuming ${\cal C}_4 \xi < 0$,  at large $A$ we obtain the asymptotic solution in the screened region,
\be
\label{nlx}
x  \simeq  - \frac{ \sqrt{ -  {2 A \xi {\cal C}_4} }}{{\cal C}_4}  \qquad (A \gg1) \;.
\ee
The scale of transition from the linear to the nonlinear regime for $\chi$ can be obtained by matching the two asymptotic solutions for $x$, eqs.~\eqref{lin_spher} and \eqref{nlx}, which gives
\be
A_V \approx \frac{2 \nu^2}{ |\xi {\cal C}_4| }  = \bigg| \frac{\mu_\Phi - 1}{2 \mu_{\Phi,2} } \bigg| \;. \label{Vainrad1}
\ee
For the the second equality, we have used the expression for $\mu_{\Phi,2}$ in eq.~\eqref{muphiexp} and that $\mu_{\Phi} - 1 = \xi^2/\nu = \mu_\chi^3 \nu^2/\xi$.
The  Vainshtein overdensity is thus set by the ratio between the linear and second-order modification of the effective Poisson equation. 

For overdensities larger than $A_V$, the corrections to the solutions for the metric fluctuations $y$ and $z$ in eq.~\eqref{metricsolns} becomes smaller than the effects of modified gravity in the linear regime in eq.~\eqref{lin_spher}. This can be seen by 
 rewriting the Vainshtein solution \eqref{Vainrad1} as
\be
x \simeq  \mu_\chi  \sqrt{A A_V} \qquad (A \gg1)\;, 
\ee
and inserting this in eqs.~\eqref{Vain1} and \eqref{Vain2}  to compute $y$ and $z$. 
 This gives
\be \label{metricsolns}
y  \simeq  A + (\mu_{\Phi}-1)  \sqrt{ {A}{A_V}}    \;, \qquad z \simeq A  + (\mu_{\Psi}-1)  \sqrt{ {A}{A_V}}   \qquad (A \gg1) \;.
\ee
This shows that, for $M^2$ constant, general relativity is recovered up to  corrections of order $\sim (\mu_{\Phi}-1) (A_V/A)^{1/2}$.
 It is instructive to compare these solutions to those obtained from eqs.~\eqref{Vain1}--\eqref{Vain3} in the perturbative regime $A \ll 1$, i.e.,
\be
x  \simeq \mu_{\chi} A \left(1  - \frac{A}{2A_V} \right) \ , \qquad y \simeq  A \left(1 + (\mu_\Phi -1) \left(1  - \frac{A}{2A_V} \right) \right)  \qquad (A\ll1) \;
\ee
(and an analogous expression for $z$). This shows that our perturbative solution \eqref{sol_NL} does not hold in the Vainshtein regime.

We can compare the Vainshtein scale to  the usual nonlinear scale for dark matter density perturbations $k_{\rm NL}$, by looking at $\langle A^2  \rangle_R$, defined as the 
mean squared fluctuations of $A$ over a sphere of radius $R$.
We assume that the matter distribution is described by the power spectrum of a scaling universe (see e.g.~\cite{Pajer:2013jj}), i.e.
\be
\langle \delta_{\vec k} \delta_{\vec k'} \rangle = (2 \pi)^3 \delta(\vec k + \vec k') P(k) \;, \qquad P(k) =  \left( \frac{2 \pi}{k_{\rm NL}} \right)^3  \left( \frac{k}{k_{\rm NL}} \right)^n \;,
\ee 
where $n\simeq -2$ near the nonlinear scale in the real universe.  In this case we have
\be
\label{AR}
\langle A^2  \rangle_{R = \frac{2\pi}{k}} =  \frac{ \pi }{3 +n}  \, {\Omega_{\rm m}^2}    \left( \frac{k}{k_{\rm NL}} \right)^{{3+n}} \;.
\ee
Evaluating this at the onset of the Vainshtein regime using eq.~\eqref{Vainrad1}, we find
\be
\frac{\kV}{\knl} \approx \left(  \frac1{\Omega_{\rm m}} \bigg| \frac{\mu_\Phi - 1}{2 \mu_{\Phi,2} } \bigg|  \sqrt{ \frac{3+n}{ \pi}} \right)^{\frac2{3+n}}  \;.
\label{Vainscale1}
\ee
For  nonlinear couplings in the effective Poisson equation larger than the linear modifications, $|\mu_{\Phi,2}| \gtrsim |\mu_\Phi - 1|$, the Vainshtein screening becomes already important  on mildly nonlinear scales, i.e.~$\kV \lesssim \knl  $, as expected.

%
%
\subsubsection{Quartic screening } 

Let us now consider  $\alphac=\alphaset=0$.
The cubic equation obtained in this case is 
\be
 (1-\alphaq) a_3 x^3 +   a_2 x^2 - \nu x + A ( \xi  - 2 a_3 x)  =0 \;,  
\ee
where
\be
 a_2 \equiv \frac12 \big(  {\cal C}_4 + 3 \xi \bone  - 3 \alphaB \alphaT \big) \;, \qquad a_3 \equiv \frac12 \big( \alphaT  - \alphaq (1+\alphaT)  \big)\;.
\ee
In the  large density region $A\gg1$, there are three solutions,
\be
x  \simeq  \pm \sqrt{\frac{2 A}{1-\alphaq}} \;, \qquad x  \simeq \frac{\xi}{2 \, a_3} \qquad (A\gg1) \;, \label{solVaincubic}
\ee
These can be matched to the $\xi A/\nu$ solution for $A \ll 1$, depending on the parameters above.
One finds \cite{Kimura:2011dc} (the symbols $\land$ and $\lor$ respectively denote the logical ``and'' and ``or'')

\begin{itemize}

\item $\alphaq < 1 \;, \quad a_3<0:$
\be
x  \simeq  \sqrt{\frac{2 A}{1-\alphaq}}  \qquad  (\xi > \xi_*   \land  \xi > 0) \;, \qquad 
 x \simeq  - \sqrt{\frac{2 A}{1-\alphaq}} \qquad (  \xi < \xi_* \land  \xi < 0)\;,
\ee
where $\xi_* \equiv {2 a_2^3}/{[9 (1-\alphaq) ( a_2^2  + 3 \tilde  a_3 /4  )]}  $ with $  \tilde a_3 \equiv 4 a_3 (1-\alphaq) \nu $.

\item $\alphaq < 1 \;, \quad a_3> 0:$
\be
x  \simeq \frac{\xi}{2\, a_3} \qquad ( -  \xi_+ <   \xi <          - \xi_- )  \; ,
\ee
where $\xi_\pm = ({a_2 \pm \sqrt{a_2^2 +  \tilde  a_3 }})/({1-\alphaq})$.

\item $\alphaq > 1 \;, \quad a_3 > 0:$
\begin{align}
x &\simeq  \frac{\xi}{2 \, a_3}   \\
( &    a_2 >  \sqrt{ -  \tilde a_3  } \land \xi < - \xi_+)  \ \lor  \ (   a_2 < -  \sqrt{ -  \tilde a_3 } \land  \xi > - \xi_-  ) \  \lor \ (  -  \sqrt{ -  \tilde a_3 } <a_2 <  \sqrt{ -  \tilde a_3 }) \nonumber \;.
\end{align}

\item $\alphaq > 1 \;, \quad a_3< 0:$
There are no solutions for this case.

\end{itemize}

Using the first two solutions in eq.~\eqref{solVaincubic} to solve for $y$ and $z$ one finds
\be
y  \simeq  \frac{A}{1-\alphaq} \pm  \xi \sqrt{ \frac{2A}{1-\alphaq}} \;, \qquad  z  \simeq   \frac{A}{1-\alphaq} \pm  \alphaB \sqrt{ \frac{2A}{1-\alphaq}}  \qquad (A \gg 1) \;. 
\ee
Thus, general relativity is recovered for $A \gg 1$ when $M^2=$ const.~and $\alphaq =0 $. 
In this case the transition between the screening and linear regime takes place at $A_V \approx 2  \nu^2 / \big( \xi^2 (1-\alphaq) \big)$.
The third solution instead gives
\be
y  =  (1+\alphaT){A} +{\cal O}(A^{1/2}) \;, \qquad  z =    A +{\cal O}(A^{1/2})  \qquad (A \gg 1) \;,
\ee
which shows that general relativity is recovered for $M^2=$ const.~and $\alphaT =0 $. In this case $A_V \approx \nu/(2 a_3)$. Using $\alpha_I$, $a_3 \simeq -\frac23 \mu_{\Phi,2}/\mu_\chi^2 $ and  $\nu = (\mu_{\Phi} - 1 -\alphaT)/\mu_\chi^2$ valid for small $\alpha_I$ and following the procedure outlined above, the Vainshtein scale reads
\be
\frac{\kV}{\knl} \approx \left( \frac{3}{4 \Omega_{\rm m}} \bigg| \frac{ \mu_\Phi - 1 - \alphaT}{\mu_{\Phi,2} }  \bigg|  \sqrt{ \frac{3+n}{\pi}} \right)^{\frac2{3+n}} \;,
\ee 
which leads to analogous conclusions as eq.~\eqref{Vainscale1} for the Vainshtein scale.

%
%
%

%
%

\section{After GW170817/GRB170817A} \label{gw17}

Recently, the association of GW170817 \cite{TheLIGOScientific:2017qsa} and GRB 170817A \cite{Goldstein:2017mmi}  events  allowed  to establish with very high precision that the speed of gravitational waves equals the one of light, $c_T=1$  \cite{Monitor:2017mdv}. To avoid tuning, the speed of gravitational waves must be unaffected not only for our particular cosmological solution  but also for nearby solutions obtained by slightly changing the matter abundance \cite{Creminelli:2017sry,Ezquiaga:2017ekz,Baker:2017hug} (see also \cite{Sakstein:2017xjx}). 

Following \cite{Creminelli:2017sry}, let us review the consequences of this result on Horndeski theories by using the nonlinear action \eqref{total_actionEFT}. The speed of propagation of tensors is affected by the operator $m_4^2$, so that we require $m_4^2 =0$. 
On the other hand, a small change of the background solution shifts $\delta g^{00}$ and $\delta K_{ij}$ by some background values, respectively denoted by $\delta g^{00}_{\rm bkgd}$ and $\delta H_{\rm bkgd} h_{ij}$.  
This  generates quadratic operators proportional to either $\delta K^\mu{}_\nu \delta K^\nu{}_\mu$ or $R$, with the net effect of shifting $m_4^2$ from $0$ to 
\be\label{51}
- \frac12   m_5^2 \delta g^{00}_{\rm bkgd} +  m_6 \delta  H_{\rm bkgd} -\frac12 (m_6 \delta g^{00}_{\rm bkgd} )^{\hbox{$\cdot$}} + m_7 \delta g^{00}_{\rm bkgd}\delta  H_{\rm bkgd} \;.
\ee
Requiring that this expression vanishes for any $\delta g^{00}_{\rm bkgd}$, $\delta \dot g^{00}_{\rm bkgd}$ and $ \delta  H_{\rm bkgd}$ implies that  \cite{Creminelli:2017sry}
\be
 m_5^2=m_6 =m_7 =0 \;,
\ee
or, using eq.~\eqref{EFTaction_masses}, 
$\alphaq=\alphac=\alphaset=0$.\footnote{The coefficients of higher order operators that we have not included in the action \eqn{total_actionEFT}, because they are not relevant in the quasi-static limit, are also constrained by requiring that $c_{\rm T} = 1$.  For example, any operator which is made up of a power $(\delta g^{00})^n$ multiplied by one of the operators proportional to $m_4^2$, $m_5^2$, $m_6$, or $m_7$ will also change the speed of tensors when the background is shifted.  Requiring that all of these operators are absent gives the constraint in terms of the Horndeski functions, which is $G_{5,X} =0$ and $2 G_{4,X} +G_{5,\phi}  =0$ for \emph{any} $X$ (see e.g.~\cite{Creminelli:2017sry}).} Since also $\alphaT=0 $, the nonlinear couplings $\mu_{\Phi,2} $, $\mu_{\Phi,22} $ and $ \mu_{\Phi,3}$ given in \eqn{muphiexp} simplify considerably and reduce to  
\be
\mu_{\Phi,2}  = \frac{\alphaM - 2 \alphaB}{2} \left( \frac{  \mu_\Phi - 1 }{\nu} \right)^{3/2} \;, \qquad
{{ \mu_{\Phi,22} }  = \frac{(\alpha_{\rm M} - 2 \alpha_{\rm B} )^2 (\mu_\Phi - 1)^2}{2 \nu^3}    }   \;,   \qquad
\mu_{\Phi,3}  =  0\;.
\ee
This means that any nonlinear effect is necessarily associated to a change in the linear solution due to either $\alphaB \neq 0$ or $\alphaM \neq 0$, i.e.~$\mu_{\Phi} \neq 1$.

{As discussed in \cite{Creminelli:2017sry}, these constraints can be relaxed if dark energy has a fixed $\dot \phi$ independent of $H$. In this case, in the EFT language we have $\delta g^{00}_{\rm back} =0$  and eq.~(\ref{51}) becomes $m_6=0$, which by eq.~(\ref{EFTaction_masses2})  implies 
$\alphac=0$.  However, $\alphaq$ and $\alphaset$ can still be non-vanishing, since there is no constraint on $m_5^2$ or $m_7$.  

 Another way out is to assume} that dark and visible matter couple to different metrics. In particular, the constraints from GW170817/GRB170817A apply in the Jordan frame of visible matter (i.e. of baryons, photons, neutrinos etc.). Denoting with a hat quantities in the dark matter frame, using $\alpha_{\rm V1}=\alpha_{\rm V2}=\alpha_{\rm V3}=0$ in the visible matter frame and eq.~(\ref{frame_tra}) one sees that $\hat \alpha_{\rm V2}$ and $\hat \alpha_{\rm V3}$ must  necessarily vanish while a non-vanishing $\hat \alpha_{\rm V1}$ can be obtained if dark matter is disformally coupled to the metric of the visible matter frame, i.e.~$\alpha_{\rm D} \neq 0$. This case, however, requires a specific treatment that takes into account the different couplings of dark and visible matter, analogously to what is done for linear perturbations in \cite{Gleyzes:2015pma}.  Additionally, since in this scenario baryons and dark matter are effectively coupled to different metrics, new non-linear interactions between baryons and dark matter would appear with respect to former non-linear treatments, as in for example \cite{Lewandowski:2014rca}.  


%
%

\section{Conclusions}
\label{concref}

{We have developed the nonlinear interactions of dark energy and modified gravity in the Effective Field Theory approach. Specifically, in Sec.~\ref{eftofdesec} we have focused on the operators describing the covariant Horndeski Lagrangians, assuming minimal coupling for matter. For these theories, we have derived all the operators up to quartic order and provided a connection between the coefficients in the Effective Field Theory of Dark Energy action and the coefficients in the covariant form of the Horndeski action, see details in App.~\ref{app:Horndeski}. 

Moreover, motivated by describing the clustering of dark matter in the presence of dark energy and modified gravity, we have focused our analysis on the quasi-static non-relativistic limit, valid on scales much smaller than the Hubble scale.  We have shown that there are only six independent operators in this limit, a result valid at any order in perturbation theory, see action in eq.~\eqref{total_actionEFT}, Sec.~\ref{nonlinear}. Three of these operators contribute only beyond linear order and introduce three new functions of time necessary to describe the nonlinear regime. The transformations of these time-dependent functions under conformal and disformal metric redefinitions are discussed in Sec.~\ref{sec:frame}.

Then, in Sec.~\ref{NMG} we have  computed the equations of motion  in  Newtonian gauge. Since in the quasi-static limit the scalar field is non-dynamical, the equations of motion are three constraints that can be used to relate the two gravitational potentials and the scalar field fluctuations to the matter density contrast. After a review of the linear case, we have computed the action and constraint equations from the nonlinear operators. We have then used these equations to compute the relation between the Newtonian potential and the matter density contrast, see eqs.~\eqref{sol_NL} and \eqref{muphiexp}. Apart from the usual linear modification in the effective Newton constant, the nonlinear couplings modify this equation beyond linear order. The new couplings in \eqn{sol_NL} are relevant for the computation of the one-loop dark matter power spectrum \cite{CLV2}. 
In order to understand the size of the effects that we have computed, in Sec.~\ref{muplots} we have presented a sampling of plots of the time-dependent parameters that are relevant for dark-matter clustering.  

In Sec.~\ref{NLreg} we have described the Vainshtein screening regime around spherical sources and the relation between the Vainshtein scale and the nonlinear scale for dark matter clustering coming from the nonlinear couplings in the continuity and Euler equations.  We have shown that the effective Poisson equation relating the Newtonian potential to the matter density contrast derived in Sec.~\ref{NMG} does not hold inside the Vainshtein regime.
Finally, we have discussed the implications of the recent gravitational wave/gamma ray burst observations on our findings and possible ways around these constraints.  

This work could be extended in several directions. For instance, it is straightforward to include in the action operators describing GLPV theories \cite{Gleyzes:2014dya,Gleyzes:2014qga} or higher-order \cite{Zumalacarregui:2013pma} degenerate theories \cite{Langlois:2015cwa,Crisostomi:2016czh} and derive analogous constraint equations.  See e.g.~\cite{Dima:2017pwp} for progresses in this direction.  Another direction is to go beyond the quasi-static approximation and include the effects of the dark-energy field's propagation.  This approach could also be used in numerical simulations, such as e.g.~\cite{Adamek:2016zes}, to parametrize in a general way the effects of modified gravity.
}

%
\section*{Acknowledgements}
The authors are pleased to thank B.~Bellazzini, E.~Bellini, B.~Bose, C.~Charmousis, P.~Creminelli, M.~Crisostomi, J.~Gleyzes, K.~Koyama, V.~Pettorino, M.~Schmidful and M.~Zumalac{\'a}rregui for many useful discussions related to this project.  M.L.~and F.V.~are also pleased to thank the workshop DARK MOD and its Paris-Saclay funding, the organizers and participants for interesting discussions.  M.L.~acknowledges financial support from the Enhanced Eurotalents fellowship, a Marie Sklodowska-Curie Actions Programme. F.V.~acknowledges financial support from ``Programme National de Cosmologie and Galaxies'' (PNCG) of CNRS/INSU, France and  the French Agence Nationale de la Recherche under Grant ANR-12-BS05-0002. The work of G.C. is supported by the Swiss National Science Foundation.

%
%

\newpage
\appendix

%
%

\section{Correspondence with the covariant theory}
\label{app:Horndeski}

\newcommand{\Gtwo}{G_2{}}
\newcommand{\Gthree}{G_3{}}
\newcommand{\Gfour}{G_4{}}
\newcommand{\Gfive}{G_5{}}
\newcommand{\Ftwo}{F_2{}}
\newcommand{\Fthree}{F_3{}}
\newcommand{\Ffour}{F_4{}}
\newcommand{\Ffive}{F_5{}}
\newcommand{\Hfour}{F_4{}}
\newcommand{\Hfive}{F_5{}}

In this appendix we provide details on the connection between the covariant Horndeski action, \eqn{Horndeski}, and the action written in the EFT language. In particular, we write the $m_I (t)$ functions (or equivalently the $\alpha_I(t)$) appearing  in the actions \eqref{quad_actionEFT} and \eqref{total_actionEFT}, in terms of the $G_I ( \phi , X)$.  While \eqn{total_actionEFT} contains only terms relevant in the quasi-static limit, here we will present all  operators up to quartic order. 

Our starting point is the covariant action, \eqn{Horndeski}, which we rewrite in unitary gauge.  
Following Ref.~\cite{Gleyzes:2013ooa}, we perform a 3+1 decomposition of the derivatives of the scalar field and of the 4-d Ricci and Einstein tensors, in terms of the extrinsic curvature  and 3-d Ricci tensors.
%
After some cumbersome but straightforward manipulations \cite{Gleyzes:2013ooa} one finds (note that $K^0_{\ 0} = K^0_{\ i} =0$)
\be
\begin{split}
S & = \int d^4x \sqrt{-g} \bigg[ \Atwo(t, \gzz) +  \Athree(t,\gzz) K  + \Bfour(t, \gzz)   \big(K^2 - K_{ij} K^{ij} \big) +\Afour  (t, \gzz) \R \\
&+  \Bfive (t,\gzz) \big( K^3 - 3 K K_{ij} K^{ij} + 2  K_{ij} K^{ik} K^j_{\ k} \big)  + \Afive (t,\gzz)   \big(  K_{ij} \R^{ij} - K R /2 \big)  \bigg]  \; ,
 \label{Horndeski_UG}
\end{split}
\ee
where  the $A_i$ and $B_i$ are generic functions of time and $\gzz$, related to the coefficients in \eqn{Horndeski} through 
\begin{equation} \label{AG}
\begin{aligned}[c]
A_2& =G_2- \sqrt{-X}\int \frac{G_{3, \phi}}{2\sqrt{-X'}} dX' \;, \\
A_3& =-\int dX' \sqrt{-X'} G_{3, X'} -2 \sqrt{-X} G_{4, \phi}\;,  \\
A_4 & =-G_4+2 X G_{4, X} +\frac{X}{2} G_{5, \phi}\;, 
\end{aligned}
\hspace{.4in}
\begin{aligned}[c]
 A_5& = -\frac{(-X)^{3/2}}{3} G_{5, X}\;, \\
B_4& =G_4+\sqrt{-X} \int dX' \frac{G_{5, \phi}}{4\sqrt{-X'}}\;, \\
B_5& =-\int dX' \sqrt{-X'} G_{5, X'} \ , 
\end{aligned}
\end{equation}
where we remind that $X=g^{00} \dot { \phi}^2 (t)$ in unitary gauge. In this appendix $\dot \phi$ is evaluated on the background. One can check that 
\be\label{rel}
A_4=-B_4 + 2 \gzz \frac{ \partial B_{4}}{\partial \, \gzz}  \,,\qquad \qquad A_5=  - \frac{1}{3} \gzz \frac{ \partial B_{5}}{ \partial \, \gzz}\,.
\ee

Next, we expand the operators in \eqn{Horndeski_UG}  as 
\begin{equation}\label{exp}
\gzz =-1+\dgzz\, , \hspace{.2in}  R^i{}_j =\delta R^i{}_j\,, \hspace{.2in} K^i{}_j =H \delta^i{}_j+\delta K^i{}_j \,.
\end{equation}
Since we are going to be comparing to the EFTofDE action which does not have a linear term proportional to $\delta K$, we first need to write the expression in \eqn{Horndeski_UG} so that it explicitly does not contain such a term, which comes from the $A_3$, $A_4$, and $A_5$ terms.  To do that, we first find the linear in $K$ terms, which are $A_3 ( t , -1 ) K$, $A_4 ( t , -1 ) H K$, and $A_5 ( t , -1 ) 6 H^2 K$.  Then we use the following identity in unitary gauge
\be
\int d^4 x \sqrt{-g} \lambda ( t ) K  = - \int d^4 x \sqrt{-g} \,  n^0 \nabla_0 \lambda (t) = - \int d^4 x \sqrt{-g} \,  \dot \lambda ( t ) \sqrt{-\gzz} \ ,
\ee
to add and subtract zero to \eqn{Horndeski_UG} in way that makes manifest that there are no $\delta K$ terms.  In particular, we add zero in the following three ways 
\begin{align}
0 & = \int d^4 x \sqrt{-g} \left( - \dot A_3 ( t , -1 ) \sqrt{-\gzz} - A_3 ( t , -1 ) K \right) \\
0 & = \int d^4 x \sqrt{-g} \left( - \partial_t \left( H A_4 ( t , -1 ) \right) \sqrt{-\gzz}  - A_4 ( t , -1 ) H K\right) \\
0 & = \int d^4 x \sqrt{-g} \left( - \partial_t \left( 6 H^2 A_5 ( t , -1 ) \right) \sqrt{-\gzz} - 6 A_5 ( t , -1) H^2 K \right) \ .
\end{align}
which allows us to rewrite the action \eqn{Horndeski_UG} as
\be
\begin{split}
S & = \int d^4x \sqrt{-g} \bigg\{ A_2 ( t , \gzz) - \sqrt{-\gzz} \, \dot A_{00} \\
& \hspace{1in} +  \left[ \Athree(t,\gzz) K - A_3 ( t , -1) K  \right]    \\
& \hspace{1in}+ \left[ \Bfour(t, \gzz)   \big(K^2 - K_{ij} K^{ij} \big)  - A_4 ( t , -1) H K \right]  \\
& \hspace{1in} +  \left[ \Bfive (t,\gzz) \big( K^3 - 3 K K_{ij} K^{ij} + 2  K_{ij} K^{ik} K^j_{\ k} \big) - 6 A_5 ( t , -1 ) H^2 K \right]  \\
& \hspace{1in} +\Afour  (t, \gzz) \R+ \Afive (t,\gzz)   \big(  K_{ij} \R^{ij} - K R /2 \big) \bigg\}    \; ,
 \label{Horndeski_UG2}
\end{split}
\ee
where 
\be
A_{00} \equiv  A_3 ( t , -1) + H A_4 ( t , -1 ) + 6 H^2 A_5 ( t , -1)   \;.
\ee
Notice that now, in this equation we have transferred all of the linear terms proportional to $\delta K$ into the $\gzz$ dependent term on the first line.

In order to most easily compare with the EFT actions in \eqn{quad_actionEFT} and \eqn{total_actionEFT}, it is useful to rewrite the ${}^{(4)}\!R$ part of the action as 
\be \label{trick1}
\int d^4 x \sqrt{-g} \frac{M_*^2 f(t)}{2} {}^{(4)}\!R = \int d^4 x \sqrt{-g} \frac{M_*^2 }{2} \left[  f(t)\left(  R + K_{ij} K^{ij} - K^2 \right) - 2 \dot f (t) \sqrt{-\gzz} K \right] \ .
\ee

Let us now expand the action \eqref{Horndeski_UG2} up to fourth order in the perturbations of \eqn{exp}, writing the $n$-th order Lagrangian as $\mathcal{L}^{(n)}$, so that $S = \int d^4 x \sqrt{-g} \sum_n \mathcal{L}^{(n)}$.  At quadratic order, in terms of the operators $\delta \mathcal{K}_2$ and $\delta \mathcal{G}_2$ defined in \eqn{K2}, we have  
\begin{align}
\label{2}
\mathcal{L}^{(2)}= - \frac{M^2 }{2}  \delta\mathcal{K}_2 +  \frac{m_2^4  }{2} \, ( \dgzz )^2 + \left( \frac{M_*^2 \dot f }{2}  - \frac{m_3^3 }{2}  \right)\, \delta K + \left( \frac{m_4^2 }{2} + \frac{ B_5^{(1,0)}}{4} \right) \dgzz \delta R + B_5 \delta\mathcal{G}_2 \,,
\end{align}
where 
\begin{equation}
\begin{split}
M^2 & =   -2 A_4  -  6 H A_5 \ , \\
m_2^4 & = A_2^{(0,2)} +3H A_3^{(0,2)} + 6 H^2 A_4^{(0,2)} + 6 H^3 A_5^{(0,2)}  + \frac{\dot A_{00}}{4} \ , \\
m_3^3 & = M_*^2 \dot f  - 2 A_3^{(0,1)} -8H A_4^{(0,1)} -12 H^2 A_5^{(0,1)} \ , \\
m_4^2 & = 2 B_4^{(0,1)} - H B_5^{(0,1)} - \half B_5^{(1,0)} \ ,
\end{split}
\end{equation}
and we have used the notation $A^{(m,n)} \equiv \partial_t^m \partial_{\gzz}^n A |_{( t , -1)}$.
First, notice that \eqn{2} is of the form of the quadratic Lagrangian \eqref{quad_actionEFT} (when ${}^{(4)}\!R$ is expanded as in \eqn{trick1}), apart from the $\delta \mathcal{G}_2$ term.  However, this term can be eliminated by using the identity  \cite{Gleyzes:2013ooa}
\be
\int d^4 x \, \sqrt{-g} \left[ \lambda(t) \left( K_{ij}R^{ij} - \half K R \right) - \half \dot \lambda ( t) \sqrt{-\gzz} R \right] = 0  \;,
\ee
and then expanding it around the FLRW background.  This eliminates both the $B^{(1,0)}_5$ and $B_5$ terms in \eqn{2} so that indeed the coefficient of $\dgzz \delta R$ is $m_4^2 / 2$, as desired.   The other coefficients given in \eqn{2} are the ones that appear in the quadratic Lagrangian in \eqn{quad_actionEFT}.  In terms of the $\alpha$ parameters in \eqn{EFTaction_masses}, we have (using \eqn{rel})
%
%
\begin{align}
\begin{split}
\alphaT & \equiv - \frac{ 2 \, m_4^2}{M^2} = \frac{2}{M^2} \left( A_4 +B_4 +3 HA_5 +\frac12  B_5^{(1,0)} \right) \;, \\
\alphaB & \equiv \frac{M_*^2 \dot f - m_3^3}{2 M^2 H} =  \frac{1}{M^2H}  \big(  A_{3}^{(0,1)} + 4 H A_{4}^{(0,1)}  +6 H^2 A_{5}^{(0,1)} \big) \;, \\
\alphaM & \equiv \frac{ M_*^2 \dot f + 2 ( m_4^2)^{\hbox{$\cdot$}}}{M^2 H} =  \frac{\partial_t(-2 A_4 -6H  A_5)  }{M^2H}  \;.  \label{alphasAB}
\end{split}
\end{align}

Now we move on to the cubic Lagrangian.  In terms of the operators $\delta \mathcal{K}_2$, $\delta \mathcal{K}_3$ and $\delta \mathcal{G}_2$ defined in eqs. (\ref{K2}) and (\ref{K3}), at third order we have
\begin{align}
\mathcal{L}^{(3)}&=\     - \frac{m_6}{3} \left(\delta\mathcal{K}_3+3 \dgzz \delta \mathcal{G}_2\right)   -\frac{m_5^2}{2} \dgzz \delta \mathcal{K}_2   \label{333}    \\
 & \hspace{.3in}   + m_{(3),1}^4 (\dgzz)^3 +\left( \frac{M_*^2 \dot f}{8} +  m_{(3),2}^2   \right) \,(\dgzz)^2 \delta K + m_{(3),d_1}^2 (\dgzz)^2 \delta R \,, \nonumber
\end{align}
with coefficients given by 
\begin{equation}
\begin{split}
m_5^2 & = -2 \left( A_4^{(0,1)} + 3 H A_5^{(0,1)} \right)   \ , \hspace{.3in} m_6  = - 3 A_5  \ , \\
m_{(3),1}^4 & = \frac{1}{6} A_2^{(0,3)} + \half H A_3^{(0,3)} + H^2 A_4^{(0,3)} +H^3 A_5^{(0,3)}  + \frac{ \dot A_{00}}{16} \ ,  \\ 
m_{(3),2}^2 & =  - \frac{M_*^2 \dot f}{8} + \half \left( A_3^{(0,2)} + 4 H A_4^{(0,2)} + 6 H^2 A_5^{(0,2)} \right)  \ ,  \\
m_{(3),d_1}^2 & = \frac{1}{4} \left( 2 B_4^{(0,2)} - H B_5^{(0,2)  }  - \frac{1}{4} B_5^{(1,0)} \right) =  \frac{1}{8} \left(m_5^2 + m_4^2 + H m_6 \right) \ ,
\end{split}
\end{equation}
where we have used \eqref{rel} in the second equality of the last line which shows that the coefficient $m_{(3),d_1}^2$ is not an independent parameter (the subscript ``$d$'' stands for dependent).
Notice that only $m_6$ and $m_5^2$ are relevant in the quasi-static limit.  The corresponding $\alpha$ parameters are 
\begin{align}
\begin{split}
\bone&  \equiv \frac{2 m_5^2+ 2 H m_6}{M^2} = - \frac{2}{M^2}  \left( 2 A_4^{(0,1)} + 6 H A_5^{(0,1) }+ 3 H A_5 \right) \ , \\
 \btwo & \equiv \frac{2 H m_6}{M^2} = -\frac{6 H A_5}{M^2} \ . 
 \end{split}
\end{align}

Finally at fourth order we have
\begin{align}
\begin{split}
\mathcal{L}^{(4)} &=   - \frac{m_7}{3} \dgzz \delta \mathcal{K}_3 + m_{(4),1}^4 (\dgzz)^4 + \left( \frac{M_*^2 \dot f }{16} + m_{(4),2}^3 \right) (\dgzz)^3 \delta K + m_{(4),3}^2 (\dgzz)^2 \delta \mathcal{K}_2 \\
& \hspace{.3in} +m_{(4),d_1}^3 (\dgzz)^3 \delta R + m_{(4),d_2} (\dgzz)^2 \delta \mathcal{G}_2  \ ,
\end{split}
\end{align}
where the coefficients are given by 
\begin{align}
m_7 & = - 3 A_5^{(0,1)} \ , \\ 
m_{(4),1}^4 & = \frac{1}{24} \left( A_2^{(0,4)} + 3 H A_3^{(0,4)} + 6 H^2 A_4^{(0,4)} + 6 H^3 A_5^{(0,4)} + \frac{15}{16} \dot A_{00} \right) \ ,\nonumber \\
m_{(4),2}^3 & = - \frac{M_*^2 \dot f}{16}  + \frac{1}{6} \left( A_3^{(0,3)} + 4 H A_4^{(0,3)} + 6 H^2 A_5^{(0,3)}  \right) \ ,  \hspace{.1in}  m_{(4),3}^2  =  \half \left( A_4^{(0,2)} + 3 H A_5^{(0,2)} \right) \ , \nonumber \\
   m_{(4),d_1}^2 &  = \frac{1}{12} \left( 2 B_4^{(0,3)} - H B_5^{(0,3)}  - \frac{3}{8} B_5^{(1,0)} \right) \ ,  \hspace{.1in} m_{(4),d_2}  = \half B_5^{(0,2)} \ . \nonumber
\end{align}
Similar to the cubic case, the coefficients $ m_{(4),d_1}^2$ and $m_{(4),d_2} $ are not independent parameters because they can be written as
\begin{align}
\begin{split}
m_{(4),d_1}^2 & = \frac{1}{48} \left( 3 m_4^2  + 3 m_5^2 + 2 H m_7 + 5 H m_6 - 8 m_{(4),3}^2    \right) \ , \\
m_{(4),d_2} & = - \frac{1}{2} \left( m_6 + m_7 \right) \ . 
\end{split}
\end{align}
Notice that only $m_7$ is relevant in the quasi-static limit, and it is related to the $\alpha$ parameters by 
\be
\bthree \equiv \frac{ 4 H m_7+ 2 H m_6}{M^2} =  - \frac{ 12 H A_5^{(0,1)} +6 H A_5}{M^2}   \ . 
\ee

Now that we have the $m_I$ parameters in terms of the $A_I$ and $B_I$, we will use \eqn{AG} to replace the $A_I$ and $B_I$ with the Horndeski $G_I$ functions to finally obtain the parameters of the EFTofDE, the $m_I$, in terms of the Horndeski $G_I$.  For the $G_I$ functions, we use the notation $ G_I^{(m,n)} \equiv \partial_\phi^m \partial_X^n G_I |_{( \phi , - \dot \phi^2 )}$, and we use $\phi$ to denote the background value.  We start with the parameters that are relevant in the quasi-static limit
\begin{align}\label{A23}
\begin{split}
M^2 & = 2 G_4 + \dot \phi^2 \left( 4  G^{(0,1)}_4 +  G_5^{(1,0)} \right) + 2 H \dot \phi^3 G_5^{(0,1)} \;, \\
m_4^2 & = \dot \phi^2 \left( 2  G_4^{(0,1)} +  G_5^{(1,0)} \right) + H \dot \phi^3 G_5^{(0,1)} - \dot \phi^2 \ddot \phi G_5^{(0,1)} \;,\\
m_3^3 & = M_*^2 \dot f - 2 \dot \phi G_4^{(1,0)} - \dot \phi^2 \left(8 H  G_4^{(0,1)} + 4 H  G_5^{(1,0)} \right) + \dot \phi^3\left( 2  G_3^{(0,1)}+4  G_4^{(1,1)} - 6 H^2  G_5^{(0,1)}   \right)\;, \\
&\hspace{.2in}  + \dot \phi^4\left( 16 H  G_4^{(0,2)} + 4 H G_5^{(1,1)} \right) + 4 H^2 \dot \phi^5 G_5^{(0,2)} \;,\\
m_5^2 & = - \dot \phi^2 \left( 2  G_4^{(0,1)} +  G_5^{(1,0)} \right) - 3 H \dot \phi^3 G_5^{(0,1)} + \dot \phi^4\left( 4  G_4^{(0,2)} +  G_5^{(1,1)} \right) + 2 H \dot \phi^5 G_5^{(0,2)} \;,\\
m_6 & = \dot \phi^3 G_5^{(0,1)} \;,\\
m_7 & = - \frac{3\dot \phi^3}{2} G_5^{(0,1)} + \dot \phi^5 G_5^{(0,2)} \ . 
\end{split}
\end{align}
The other parameters are given by 
\begin{align}
m_2^4 & = \frac{\dot  A_{11}}{4}  + \frac{3}{2} H \dot \phi G_4^{(1,0)}  + \frac{\dot \phi^2}{2} \left( - G_4^{(2,0)} + \ddot \phi G_3^{(0,1)}   + 2 \ddot \phi G_4^{(1,1)} \right)  - \half \ddot \phi \gfouroz \nonumber \\
& \hspace{.0in} + \frac{\dot \phi^3}{2} { H} \left( 3  G_3^{(0,1)} + 12  G_4^{(1,1)} - 3 H^2 G_5^{(0,1)} \right) + \dot \phi^4 \left( \gtwozt - \half \gthreeoo + 18 H^2 \gfourzt + 6 H^2 \gfiveoo \right) \nonumber \\
& \hspace{.0in} - 3 \dot \phi^5 { H }\left(    \gthreezt + 2  \gfourot - 2 H^2 \gfivezt \right) -  3 \dot \phi^6 { H^2} \left( 4  \gfourzth +   \gfiveot \right) -2 \dot \phi^7 H^3 \gfivezth \;,  
\end{align}
\begin{align}
m_{(3),1}^4 & = \frac{ \dot A_{11}}{16} + \frac{3}{8} \dot \phi H \gfouroz + \frac{\dot \phi^2}{8} \left( - \gfourtz + \ddot \phi \left( \gthreezo + 2 \gfouroo \right) \right)  - \frac{1}{8} \ddot \phi \gfouroz  \nonumber \\
& \hspace{.0in} + \frac{\dot \phi^3 }{8} { H} \left(  \gthreezo + 6  \gfouroo - H^2 \gfivezo \right)  + \frac{\dot \phi^4}{24} \gthreeoo  +  \frac{\dot\phi^5}{4}  { H}\left( 2   \gthreezt + 6  \gfourot - 3 H^2 \gfivezt \right)  \nonumber \\
& \hspace{.0in}  + \frac{ \dot \phi^6}{12} { \left( { 2 \gtwozth} - { \gthreeot} + 60 H^2 \gfourzth + {18H^2 \gfiveot} \right)   -  \frac{\dot \phi^7  }{2} H \left( { \gthreezth} +  2 \gfouroth - { 3 H^2 \gfivezth} \right) } \nonumber \\
& \hspace{.0in} - \dot \phi^8 { H^2} \left(  2  \gfourzf +  \frac12  \gfiveoth  \right) - \frac{\dot \phi^9}{3} H^3 \gfivezf    \;,
\end{align}
\begin{align}
m_{(3),2}^2& =  - \frac{M_*^2 \dot f}{8} + \frac{ \dot \phi \gfouroz}{4} + { \frac{\dot \phi^3}{4} \left( {\gthreezo}  + 4 \gfouroo - { 3 H^2 \gfivezo} \right) + 2 \dot \phi^4 H \left( 3 \gfourzt +   \gfiveoo \right) } \nonumber \\
& \hspace{.0in}  - \frac{\dot \phi^5}{2} \left( {\gthreezt} + 2 \gfourot - 6 H^2 \gfivezt \right) - \dot \phi^6 \left(  4 H \gfourzth + H \gfiveot \right) - \dot \phi^7 H^2 \gfivezth \;, 
\end{align}
\begin{align}
m_{(4),1}^4 & =\frac{5 \dot A_{11}}{128} + \frac{15}{64} H \dot \phi \gfouroz + \frac{\dot \phi^2}{64} \left( - 5 \gfourtz + \ddot \phi \left( 5 \gthreezo + 10 \gfouroo \right) \right)  \nonumber \\
& \hspace{.0in}  + \frac{3}{64}  \dot \phi^3 H \left(  \gthreezo + 8  \gfouroo -  H^2 \gfivezo \right) + \frac{5 \dot \phi^4}{192} \gthreeoo  \nonumber \\
& \hspace{.0in} + \frac{\dot \phi^5}{32} { H} \left( {3}  \gthreezt + 12  \gfourot - 4 {H^2 } \gfivezt \right)  + \frac{\dot \phi^6}{96} \gthreeot  - \frac{5}{64} \ddot \phi \gfouroz \nonumber \\
& \hspace{.0in} +  \frac{\dot \phi^7}{16}  H \left( {3}  \gthreezth +  8 \gfouroth - 6 H^2 \gfivezth \right)   + \frac{\dot \phi^8}{48} \left( 2 {\gtwozf} -   \gthreeoth  + 84 H^2 \gfourzf + 24 H^2 \gfiveoth \right)     \nonumber \\
& \hspace{.0in} - \frac{\dot \phi^9}{8} H \left( { \gthreezf} +2  {  \gfourof} -4  {H^2 \gfivezf} \right)   - \frac{\dot \phi^{10}}{8} H^2  \left( 4   G_4^{(0,5)} +G_5^{(1,4)} \right) - \frac{ \dot \phi^{11}  }{12} H^3  G_5^{(0,5)} \;, 
\end{align}
\begin{align}
m_{(4),2}^3 & = - \frac{M_*^2 \dot f }{16} + {  \frac{ \dot \phi }{8}\gfouroz +  \frac{\dot \phi^3}{24} \left( {\gthreezo} + 6 { \gfouroo} - 3 {H^2 \gfivezo} \right) + \frac{\dot \phi^5}{12} \left( 2 \gthreezt + 6 { \gfourot} - {9 H^2 \gfivezt} \right) } \nonumber \\
& \hspace{.0in} + \frac{ \dot \phi^6}{3} H \left( {10  \gfourzth} + 3 \gfiveot \right)   - \frac{\dot \phi^7}{6} \left(  {\gthreezth} + 2 { \gfouroth} - {9 H^2 \gfivezth} \right) \nonumber \\
& \hspace{.0in}  - \frac{\dot \phi^8}{3} H \left( {4  \gfourzf} + {  \gfiveoth} \right)  - \frac{ \dot \phi^9 H^2 }{3} \gfivezf \;, 
\end{align}
\begin{align}m_{(4),3}^2 & =  - \frac{3 H \dot \phi^3 }{8}\gfivezo + \frac{ \dot \phi^4 }{2} \left( 3 \gfourzt + \gfiveoo \right) + \frac{3 H \dot \phi^5 }{2} \gfivezt - { \frac{\dot \phi^6}{4} \left( 4 \gfourzth + {\gfiveot} \right)} - \half H \dot \phi^7 \gfivezth  \ .
\end{align}
We have defined the function $A_{11} ( t ) = H A_4 ( t , -1) + 6 H^2 A_5 ( t , -1)$ for convenience.  The time derivative that appears in the above equations is given explicitly as 
\begin{align}
\dot A_{11} & = - \dot H G_4 - \dot \phi { H \left(  \gfouroz   +  \ddot \phi \left( 2\gfourzo + \gfiveoz \right) \right)}   -\dot \phi^2 \left( 2 \dot H \gfourzo  + \half \dot H \gfiveoz + 6 H^2 \ddot \phi \gfivezo  \right)   \nonumber \\
&  \hspace{.0in} - \frac{\dot \phi^3}{2}  H \left( 4  \gfouroo +   \gfivetz + 8  \dot H \gfivezo - 2 \ddot \phi \left( 4 \gfourzt + \gfiveoo \right) \right) -2  \dot \phi^4 H^2 \left(   \gfiveoo -2   \ddot \phi \gfivezt \right) \ . 
\end{align}

%
%


\section{Stueckelberg trick up to second order}
\label{Stuek}

In this section we  derive some useful relations to restore the general covariance of the action 
and write it in a generic coordinate system. In unitary gauge, the action is constructed by writing all of the operators in terms of the metric that are invariant under time-dependent spatial diffeomorphisms $x^i \rightarrow x^i - \xi^i ( \xvec , t)$.  This action has one extra scalar degree of freedom, associated with the breaking of time diffeomorphisms, which can be explicitly introduced using the Stueckelberg trick.  To do this, one first performs a broken time diffeomorphism on the action: $t \rightarrow \tilde t = t - \xi^0 ( \xvec, t)$.  Then one makes the replacement $\xi^0 (x ( \tilde x)) \rightarrow - \tilde \pi ( \tilde x)$, where $\pi$ is the Goldstone boson that nonlinearly realizes the time diffeomorphism symmetry, which is restored if $\pi$ transforms like $\pi ( \xvec , t ) \rightarrow \pi( \xvec, t) + \xi^0( \xvec , t)$.  Finally, one writes the integral in the action as being over $\tilde x$ instead of $x$.

In the end, this has the effect that one can introduce the scalar by performing \cite{ArkaniHamed:2003uy,Creminelli:2006xe,Cheung:2007st}
\beq
t \to t + \pi (t, \vec x) \; ,
\eeq
in the action.  Specifically, one can simply replace any function of time $f(t)$ up to third order with 
\begin{align}
f &\to f + \dot f \pi    + \frac12 \ddot f \pi^2 + \frac16 f^{(3)} \pi^3+ {\cal O} (\pi^4)\; , \label{trans_ST_6} 
\end{align}
and the metric with
\be
g^{\mu \nu} \to (\delta^\mu_\alpha + \delta^\mu_0 \partial_\alpha \pi ) (\delta^\nu_\beta + \delta^\nu_0 \partial_\beta \pi ) g^{\alpha \beta} \; . 
\label{diff_trans}
\ee
In particular, the metric component $g^{00}$ transforms exactly as
\beq
g^{00} \to g^{00} +  2 g^{0 \mu} \partial_\mu \pi + g^{\mu \nu} \partial_\mu \pi \partial_\nu \pi \;.
\eeq

For the other perturbed geometric quantities, we only need their change at quadratic order in the perturbations.  For those quantities, it is useful to write the metric in ADM form, i.e.~as
\be \label{admmetric}
ds^2 = - N^2 dt^2 + h_{ij} (N^i dt + dx^i) (N^j dt + dx^j) \;,
\ee
where $N$ is the lapse, $N^i$ is the shift, and $h_{ij}$ is the spatial metric.  
Then, using the usual ADM metric relations, i.e.
\be
g^{00} = - \frac{1}{N^2} \;, \qquad g^{0i} = \frac{N^i}{N^2}  \;, \qquad h^{ij} = g^{ij} + \frac{N^i N^j}{N^2} \;,
\ee
\be
g_{00} = -N^2 + h_{ij} N^i N^j \; , \qquad N_i = h_{ij} N^j \; , \qquad h_{ij} = g_{ij} \ , 
\ee
and combined with the transformation eq.~\eqref{diff_trans}, we have
\begin{align}
N & \to N \left(1 - \dot \pi + \dot \pi^2 + N^i \partial_i \pi + \frac{1}{2} N^2 h^{ij}  \partial_i \pi \partial_i \pi \right) +{\cal O}(3) \;, \\
N^i & \to N^i(1-\dot \pi) + {( 1 - 2 \dot \pi) }N^2 h^{i k} \partial_k \pi +{\cal O}(3) \;, \\
h_{ij} & \to h_{ij} - N_i \partial_j \pi - N_j \partial_i \pi -  N^2 \partial_i \pi \partial_j \pi +{\cal O}(3) \;.
\end{align}
Other useful relations are the transformation for the derivatives,
\be
\partial_0 \to (1-\dot \pi + \dot \pi^2) \partial_0 +{\cal O}(3) \;, \qquad \partial_i \to  \partial_i - (1 - \dot \pi) \partial_i \pi \partial_0 +{\cal O}(3)\;.
\ee
Making use of the relations above, we can derive the transformations of  the extrinsic curvature and of the Ricci scalar and curvature from their definitions. We find
\begin{align}
\delta K \to & \ \delta K  - 3 \left(\dot H \pi +\frac12 \ddot H \pi^2 \right) - (1-\dot \pi) N h^{ij} \partial_i \partial_j \pi  + \half a^{-4} \partial_k \pi \left( - \partial_k h_{ii} + 2 \partial_{i} h_{k i}  \right)  \nonumber \\
& + \frac{H}{2 a^2 } (\partial_i \pi)^2    +   \frac{2}{a^2}  \partial_i \pi \partial_i \dot \pi  -   \frac{2}{a^2} \partial_i N \partial_i \pi  +{\cal O} (3) \;, \\ 
\delta K_{\ j}^i \to & \ \delta K_{\ j}^i - \left(\dot H \pi +\frac12 \ddot H \pi^2 \right) \delta^i_{j} - (1-\dot \pi) N h^{ik}  \partial_k \partial_j \pi  + \half a^{-4} \partial_k \pi \left( - \partial_k h_{ij} + \partial_{i} h_{k j} + \partial_{j} h_{k i} \right)  \nonumber \\
&   + \frac{H}{2a^2} (\partial_k \pi)^2  \delta^i_j  - \frac{  H}{a^2} \partial_i \pi \partial_j \pi +  \frac2{a^2} \partial_{(i} \dot \pi \partial_{j)} \pi  - \frac2{a^2} \partial_{(i} N \partial_{j)} \pi    + {\cal O} ( 3) \;, \\
R \to & \ R - 2 (1- \dot \pi) \dot h^{ij} \partial_i \partial_j \pi -\frac2{a^2} \big[ 4 H \partial_k \pi \partial_k \dot \pi + (H^2 + 2 \dot H) (\partial_k \pi)^2  \big] \nonumber  \\
& +\frac{1}{a^4} \bigg[ H(  \partial_k  h_{ii}  - \partial_i   h_{ik}) +   \partial_k \dot h_{ii} - \partial_i \dot h_{ik}\bigg] \partial_k \pi  \nonumber \\ 
& - \frac1{a^4} \PP_2 [ \pi ,\pi ] - \frac2{a^2} \PP_2 [ \pi ,\partial^{-2} \partial_i N^i ]   + {\cal O} (3) \;, \label{variationR} \\
R_{ij} \to & \ R_{ij}  + H (\partial_i \partial_j \pi + \delta_{ij} \partial^2 \pi)  + {\cal O} (2) \; , 
\end{align}
where 
\be
\mathcal{P}_2 [ \varphi_a , \varphi_b ] = \varepsilon^{ikm} \varepsilon^{jlm} \partial_ i \partial_j \varphi_a  \partial_k \partial_l \varphi_b \ .
\ee
Although the metric appears through $N$, $N^i$, and $h_{ij}$, the expressions above are only valid to second order in perturbations.  Note that we only computed the Stueckelberg transformation of $R_{ij}$ to first order as we do not need higher orders.


\newpage

 \bibliographystyle{utphys}
\bibliography{EFT_DE_biblio3}

\end{document}